\documentclass[twocolumn,showpacs,preprintnumbers,nofootinbib,prd,
superscriptaddress,10pt]{revtex4-1}

\usepackage{lmodern}
\usepackage{amsmath,amssymb}
\usepackage[normalem]{ulem}
\usepackage{textcomp}
\usepackage{hyperref}
\usepackage{bm}
\usepackage{graphicx}
\usepackage{psfrag}
\usepackage[usenames,dvipsnames]{xcolor}
\usepackage[utf8]{inputenc}
\usepackage{lipsum}
\usepackage{aas_macros}
\usepackage{booktabs}
\usepackage{array}

\hypersetup{
    bookmarks=true,         % show bookmarks bar?
    unicode=false,          % non-Latin characters in Acrobat???s bookmarks
    pdftoolbar=true,        % show Acrobat???s toolbar?
    pdfmenubar=true,        % show Acrobat???s menu?
    pdffitwindow=false,     % window fit to page when opened
    pdfstartview={FitH},    % fits the width of the page to the window
	pdftitle={An improved IMR model for BBHs on elliptical orbits},
    pdfauthor={Pratul Manna et al.},
    pdfsubject={Gravitational Waves},   % subject of the document
    pdfkeywords={gravitational waves, eccentric compact binaries, general 
		relativity},
    pdfnewwindow=true,      % links in new window
    colorlinks=true,       % false: boxed links; true: colored links
    linkcolor=red,          % color of internal links
    citecolor=Blue,        % color of links to bibliography
    filecolor=magenta,      % color of file links
    urlcolor=Blue,           % color of external links
    linktocpage=true
}

\allowdisplaybreaks[4]

\DeclareFontFamily{OT1}{pzc}{}
\DeclareFontShape{OT1}{pzc}{m}{it}{<-> s * [1.10] pzcmi7t}{}
\DeclareMathAlphabet{\mathpzc}{OT1}{pzc}{m}{it}

\newcommand{\hlm}{\mathpzc{h}_{\ell m}}

\newcommand{\lm}{_{\ell m}}

\graphicspath{{./figures/}}

\newcommand{\cmmnt}[1]{}

\newcolumntype{C}[1]{>{\centering\arraybackslash}p{#1}}

\begin{document}

\title{Improved inspiral-merger-ringdown model for BBHs on elliptical orbits}
\date{\today}

\author{Pratul Manna}
\email{dipakpratul2014@gmail.com}
\affiliation{Department of Physics, Indian Institute of Technology Madras, Chennai 600036, India}
%\affiliation{Centre for Strings, Gravitation and Cosmology, Department of Physics, Indian Institute of Technology Madras, Chennai 600036, India}
\author{Tamal RoyChowdhury}% \orcidlink{0000-0003-4242-0532}}
\email{tamal@uwm.edu}
\affiliation{University of Wisconsin Milwaukee, Milwaukee, Wisconsin 53201, USA}
%\affiliation{{Center for Gravitation, Cosmology and Astrophysics, University of Wisconsin Milwaukee, Milwaukee, WI 53201, USA}}
\author{Chandra Kant Mishra}
\email{ckm@physics.iitm.ac.in}
\affiliation{Department of Physics, Indian Institute of Technology Madras, Chennai 600036, India}
\affiliation{Centre for Strings, Gravitation and Cosmology, Department of Physics, Indian Institute of Technology Madras, Chennai 600036, India}

%%%%%%%%%%%%%%%%%%%%%%%%%%%%%%%%%%%%%%%%%
\begin{abstract}
  Gravitational waveforms capturing binary evolution through the early-inspiral phase play a critical role in extracting orbital features that nearly disappear during the late-inspiral and subsequent merger phase due to radiation reaction forces; for instance, the effect of orbital eccentricity. Phenomenological approaches that model compact binary mergers rely heavily on combining inputs from both analytical and numerical approaches to reduce the computational cost of generating templates for data analysis purposes. In a recent work, \emph{Chattaraj et al., Phys. Rev. D 106, 124008 (2022)}~\cite{Chattaraj:2022tay} constructed a dominant ($\ell=2$, $|m|=2$) mode model for nonspinning binary black holes (BBHs) on elliptical orbits. The model was constructed in time domain and is fully analytical. The current work is an attempt to improve this model by making a few important changes in our approach. The most significant of those involves identifying initial values of orbital parameters with which the inspiral part of the model is evolved. While the ingredients remain the same as in the previous work, the resulting (new) model, when compared against a set of target waveforms constructed here, produces match values better than 96.5\% for systems heavier than $80M_\odot$, while with the old model this limit on the total mass is $115M_\odot$. The updated model is validated against an independent eccentric waveform family (\textsc{TEOBResumS-Dali})
  for an initial eccentricity ($e_0$), mass ratio ($q$) and mean anomaly ($l_0$) in the range $0\lesssim e_0\lesssim0.3$, $1\lesssim q\lesssim3$ and $-\pi\leq l_0\leq\pi$, respectively. Further,
  an alternate model including the effect of higher order modes is also provided. Finally, while our model assumes nonspinning  components, we show that it could also be used for systems with component spin vectors (anti-) aligned w.r.t. the orbital angular momentum and small spin magnitudes.
\end{abstract}

\maketitle
%\tableofcontents
\section{Introduction}
\label{sec:intro}
\vskip 5pt

\vskip 5pt

\noindent Since the gravitational wave (GW) discovery event, GW150914~\cite{Abbott:2016blz}, the LIGO-Virgo-KAGRA collaboration has reported nearly 100 compact binary mergers observed during the first three observing runs 
\cite{LIGOScientific:2018mvr, LIGOScientific:2020ibl, LIGOScientific:2021usb, LIGOScientific:2021djp}.
 These numbers have doubled since
and the list continues to be dominated by signals identified as mergers of black holes in a binary; see for instance, the Gravitational Wave Open Science Center page~\cite{LSC:GWTC-GWOSC} that lists all reported events. 
While these (now also routine) observations continue to help improve our understanding of compact binary physics and astrophysics, their origins remain unknown~\cite{TheLIGOScientific:2016htt, Abbott:2020mjq, LIGOScientific:2023lpe}. Astrophysical environments where a binary is formed and processes through which it is formed may leave imprints on a binary's mass and spin parameters. However, the available statistics needs to 
grow in order to make inferences concerning the binary's origin based on mass and spin measurements alone~\cite{LIGOScientific:2016lio}. Eccentricity, on the other hand, can be a unique tool to identify a binary's origins, as dynamically formed binaries may still retain residual orbital eccentricities~\cite{Zevin:2018kzq, Ramos-Buades:2020eju, Fumagalli:2024gko} 
when observed in ground-based detectors currently operating.
\vspace{5pt}

Current template-based search pipelines make use of circular templates due to the expected circularization of most compact binary orbits caused by radiation reaction forces~\cite{PhysRev.136.B1224} as they enter the sensitivity bands of ground-based detectors such as Laser Interferometer Gravitational Wave Observatory (LIGO)~\cite{TheLIGOScientific:2014jea} and Virgo~\cite{Acernese:2015gua}. However, binaries formed through the dynamical interactions in dense stellar environments 
are likely to be observed with residual eccentricities $e_{\rm 20Hz}\sim0.1$ \cite{LIGOScientific:2019dag, Zevin:2018kzq, LIGOScientific:2023lpe}.
In fact, the first binary merger event involving an intermediate mass black hole, GW190521~\cite{Abbott:2020tfl}, is likely an eccentric merger \cite{Abbott:2020mjq, Gamba:2021gap} (see also Refs.~\cite{Kimball:2020qyd, Romero-Shaw:2021ual, OShea:2021faf, Romero-Shaw:2022xko, Iglesias:2022xfc, Gupte:2024jfe} discussing events with eccentric signatures).\footnote{Although, there are studies (see for instance, Refs.~\cite{Gupte:2024jfe, Iglesias:2022xfc, Ramos-Buades:2023yhy, Gamboa:2024hli}) that do not find clear signs of eccentricity, likely due to the short duration of the event.} While quasicircular templates should be able to detect systems with initial eccentricities $e_{\rm 10Hz} \lesssim 0.1$, binaries with larger eccentricities would require constructing templates including the effect of eccentricity\,\cite{Brown:2009ng, Huerta:2013qb}. Moreover, the presence of even smaller eccentricities ($e_{\rm 10Hz} \sim$  0.01 - 0.05) can induce significant systematic biases in extracting the source properties~\cite{Favata:2021vhw, Abbott:2016wiq, Divyajyoti:2023rht}). Furthermore, next generation ground-based detectors, Cosmic Explorer \cite{McClelland:T1500290-v3, Dwyer:2014fpa, Evans:2016mbw} and Einstein Telescope \cite{Punturo:2010zz, Hild:2010id}, due to their low frequency sensitivities, should frequently observe systems with detectable eccentricities \cite{Lower:2018seu, Tibrewal-etal-2021}.\\%

Even though inspiral waveforms from eccentric binary mergers involving nonspinning compact components are available to high post-Newtonian (PN) orders (for instance, Refs. \cite{Moore:2016qxz,Ebersold:2019kdc} provide waveforms up to 3PN order; see also Refs.~\cite{Mishra:2015bqa,  Tanay:2016zog, Boetzel:2019nfw,  Konigsdorffer:2006zt, Moore:2019xkm}), inspiral-merger-ringdown (IMR) eccentric models are less developed compared to their quasicircular counterparts.\footnote{For instance, most  (IMR) waveform models that include eccentricity are not calibrated to eccentric NR simulations and assume postinspiral circularization. Note however, the efforts of Refs.~\cite{Carullo:2023kvj, Carullo:2024smg} which model merger and ringdown stages for highly eccentric binaries that may not circularize before merger.} Numerous efforts toward constructing eccentric IMR waveforms, useful for data analysis purposes, are underway \,\cite{Hinder:2017sxy, Huerta:2016rwp, Chen:2020lzc, Setyawati:2021gom, Chattaraj:2022tay, Gamba:2024cvy}. However, these efforts do not include important physical effects such as spins (pointing along or away from a binary's orbital angular momentum) or higher order modes. Dominant mode (or quadrupole mode) models for eccentric binary black holes (BBHs) with component spins (anti-)aligned with respect to the binary's orbital angular momentum were recently developed in Refs.~\cite{Ramos-Buades:2019uvh, Chiaramello:2020ehz, Albertini:2023aol, Nagar:2022fep, Nagar:2024dzj, Nagar:2024oyk}. Since most mergers observed so far are consistent with a zero-effective spin
~\cite{LIGOScientific:2018jsj, LIGOScientific:2020kqk, KAGRA:2021duu}, 
models neglecting spin effects can still be useful~\cite{Huerta:2016rwp}.\footnote{Reference~\cite{OShea:2021faf} explores correlations between the binary's spins and eccentricity.} In addition, modeling of higher order modes also seems necessary as Refs.~\cite{Rebei:2018lzh, Chattaraj:2022tay} argue.
Eccentric versions of the effective-one-body waveforms including higher modes (HMs) \cite{Ramos-Buades:2021adz, Nagar:2021gss} and an eccentric numerical relativity (NR) surrogate model \cite{Islam:2021mha, Islam:2024rhm} also became available in the past couple of years. Alternatively, 
unmodeled search methods with little or no dependence on the signal model being searched, may be used for detecting an eccentric merger \cite{Klimenko:2005xv, Salemi:2019uea, Tiwari:2015gal}. 
Although, these methods are sensitive to high mass searches (typically $\gtrsim 70M_{\odot}$) \cite{ LIGOScientific:2019dag,LIGOScientific:2023lpe},
while most observed events have a mass smaller than this limit \cite{LIGOScientific:2020kqk, LIGOScientific:2021usb}; see, for instance, Fig. 3 of Ref.~\cite{Divyajyoti:2021uty}.\\

The present work follows our first paper \cite{Chattaraj:2022tay} and can be viewed as an update to the same; referred to as Paper I here onward. In Paper I, construction of hybrid waveforms by combining PN waveforms with NR simulations through a least-squares minimization was demonstrated. These hybrids were then used as target models to produce a fully analytical dominant (or quadrupole) mode model by matching an eccentric PN inspiral with a quasicircular merger-ringdown waveform. Subsequently, the performance of the model was checked both against the target hybrids used in training the model as well as against an independent family of eccentric waveforms (\textsc{ENIGMA}~\cite{Chen:2020lzc}). It was shown that matches between target hybrids and the model significantly improved compared to those against a circular template, at least at the low end of the binary masses and for small eccentricities considered there (see for instance, Fig. 9 of Paper I).
The current work aims to improve the model presented in Paper I in the view of efforts such as those of Refs.~\cite{Shaikh:2023ypz, Ramos-Buades:2022lgf}.
References~\cite{Shaikh:2023ypz, Ramos-Buades:2022lgf} develop a standard, gauge-independent prescription for defining orbital eccentricity using a suitable combination of dominant mode GW frequency. It should be noted that, in general relativity (GR), eccentricity is not uniquely defined (see, for instance Ref.~\cite{Arun:2004ff}) although at the leading (Newtonian) order there is consensus. The definition of Refs.~\cite{Shaikh:2023ypz,Ramos-Buades:2022lgf} reduces to the Newtonian value in both the small and large eccentricity limit and is also independent of gauge ambiguities. This motivates us to employ this definition of eccentricity in our model and investigate the improvements in its performance.\\

%%%%%%%
\subsection{Gauge invariant definition of eccentricity}
\label{sec:pn_nr_inputs}

Eccentricity is not uniquely defined in GR and thus templates computed within the framework of GR may have forms different from the observed data. Both perturbative and numerical solutions describing the compact binary dynamics use gauge-dependent constructs including the very definitions of orbital parameters that are evolved. These choices are almost never identical in any two approaches which naturally leads to inconsistencies between different models. To get rid of the ambiguity associated with the definition of eccentricity, Refs.~\cite{Ramos-Buades:2022lgf, Shaikh:2023ypz} proposed a new definition of eccentricity based on the GW frequency data. At 0PN order, this definition exactly reduces to the Newtonian definition of eccentricity.
We reproduce the necessary relations below. This new  eccentricity for an observed GW ($e_{\rm gw}$) signal is defined as 
\begin{equation}
    e_{\rm gw} = \cos(\psi/3)-\sqrt{3} \sin(\psi/3)\,,
\end{equation}

\noindent with, 
\begin{equation}
    \psi = \arctan \left(\frac{1-e^2_{\rm \omega_{22}}}{2e_{\rm \omega_{\rm 22}}}\right)\,,
\label{Gauge-inv ecc}
\end{equation}
% where,
\begin{equation}
    e_{\rm \omega_{\rm 22}} = \frac{\sqrt{\omega^{\rm p}_{\rm 22}}-\sqrt{\omega^{\rm a}_{\rm 22}}}{\sqrt{\omega^{\rm p}_{\rm 22}}+\sqrt{\omega^{\rm a}_{\rm 22}}}\,,
\label{Gauge_inv defn}
\end{equation}
where, $\omega^{\rm p}_{\rm 22}$ and $\omega^{\rm a}_{\rm 22}$ refer to the $(\ell=2, |m|=2)$ mode periastron and apastron frequencies, respectively and are functions of time. Since this new eccentricity is written in terms of frequencies that can be measured, it is also free from any gauge ambiguities. This definition is employed in the {\fontfamily{qcr}\selectfont{GW\_ECCENTRICITY}} package provided by Ref. \cite{Shaikh:2023ypz}.
%%%%%%%

\subsection{Summary of the current work}
\label{sec:summary}
Considerations related to a change in the definition of eccentricity such as the ones proposed by Refs.~\cite{Shaikh:2023ypz, Ramos-Buades:2022lgf} demand revisiting each element of model construction presented in Paper I which we intend to closely follow here. We start by comparing PN  waveforms (amplitudes from Refs.~\cite{Mishra:2015bqa,Boetzel:2019nfw,Ebersold:2019kdc} and phase from Ref.~\cite{Tanay:2016zog}) and NR simulations \cite{Hinder:2017sxy} used in constructing the hybrids in Paper I. The {\fontfamily{qcr}\selectfont{GW\_ECCENTRICITY}} package is used to identify a set of reference values for orbital eccentricity ($e_{\rm ref}$), mean anomaly ($l_{\rm ref}$) and GW frequency ($f_{\rm ref}$) for a given waveform.  
The PN model is evolved and compared with NR simulations in a time window returning maximum overlap between the two for a given set of reference values obtained using the package.\footnote{Throughout the paper, the term `overlap' is synonymous with match maximized over an initial time and phase.} The two then are matched in this window following the method of Refs.~\cite{Varma:2016dnf,Chattaraj:2022tay}. The resulting waveforms are the hybrids. Table~\ref{tab:nrsims} lists all hybrids constructed in Sec.~\ref{sec:hybrid_waveforms} along with starting values of eccentricity ($e_0$), mean anomaly ($l_0$)  and a frequency-dependent PN parameter ($x_0$) which is related to the orbit averaged GW frequency of the dominant mode $(f_0)$ via $x_{0}=(\pi M f_{0})^{2/3}$. Figure~\ref{fig:hybridplot} plots one of these hybrids together with the corresponding NR simulation. 
Finally, in Sec.~\ref{sec:22-mode-model}, a fully analytical dominant mode model is obtained by matching an eccentric PN inspiral (\textsc{EccentricTD})  with a quasicircular merger-ringdown model (\textsc{SEOBNRv5}).  Figure~\ref{fig:td-model-q123-amp} plots the model using the initial set of parameters for three representative hybrids listed in Table~\ref{tab:nrsims} providing a visual proof of closeness of the model with target hybrids. The model is subsequently validated against hybrid waveforms that are not used in training the model  as well as against an independent family of eccentric waveforms, \textsc{TEOBResumS-Dali}~\cite{Nagar:2024oyk, Placidi:2021rkh, Albanesi:2022xge, Placidi:2023ofj}. These are shown in Figs.~\ref{fig:match_ecc_temp_new} and \ref{fig:match_ecc_22TEOB}. For comparisons with \textsc{TEOBResumS-Dali}, however, we do not use values of orbital parameters associated with the hybrids but rather use randomly sampled values for eccentricity, mass ratio and mean anomaly in the range $0\lesssim e_0\lesssim0.3$, $1\lesssim q\lesssim3$, and $-\pi\leq l_0\leq\pi$. The range is chosen to match the parameter space spanned by the training set hybrids, except for orbital eccentricity, which is conservatively chosen to have a maximum value of $e_0=0.3$ and also explores near circular cases. We find that the model correctly (and gradually) reproduces the circular limit of the \textsc{TEOBResumS-Dali} model.\\ 

We also present an alternate model to the one constructed in Sec.~\ref{sec:22-mode-model}. The primary motivation here is to use PN input waveforms of higher PN accuracy ~\cite{Mishra:2015bqa,Boetzel:2019nfw,Ebersold:2019kdc, Moore:2016qxz} (compared to \textsc{EccentricTD}~\cite{Tanay:2016zog} used in constructing the model in Sec.~\ref{sec:22-mode-model}) to maximize the overlaps with target models, although this alternate model is less sensitive to high eccentricity cases due to the absence of eccentricity corrections beyond the leading order in the orbital phase. Comparisons with the hybrids and eccentric \textsc{TEOBResumS-Dali}~\cite{Nagar:2024oyk} waveforms show that such a model may provide a suitable alternative to the model constructed in Sec.~\ref{sec:22-mode-model}.
Figure \ref{fig:TT2_TEOB_optimized} displays this comparison. It is interesting to note that the mismatches seem to have visibly improved compared to those with the model based on \textsc{EccentricTD} for small eccentricity cases. This is likely due to higher PN accuracy of the amplitude and phase used in constructing the alternate model and thus matches better with the waveform \textsc{TEOBResumS-Dali}~\cite{Nagar:2024oyk}. For larger eccentricities, the performance of the two seem similar (see a discussion in Sec.~\ref{sec:TaylorT2 model}). Further, we extend this alternate model to include selected $\ell=|m|$ and $\ell-1=|m|$ modes and validate it against the HM hybrids and HM version of \textsc{TEOBResumS-Dali}~\cite{Nagar:2024oyk}. Note that, these are precisely the modes that are included in \textsc{SEOBNRv5HM}~\cite{Pompili:2023tna}, the quasicircular model used for the merger-ringdown part, and included in our hybrids; see Fig.~\ref{fig:hybridplot}.\\

\begin{figure*}[t!]
    \centering
    
    \includegraphics[width=0.49\textwidth]{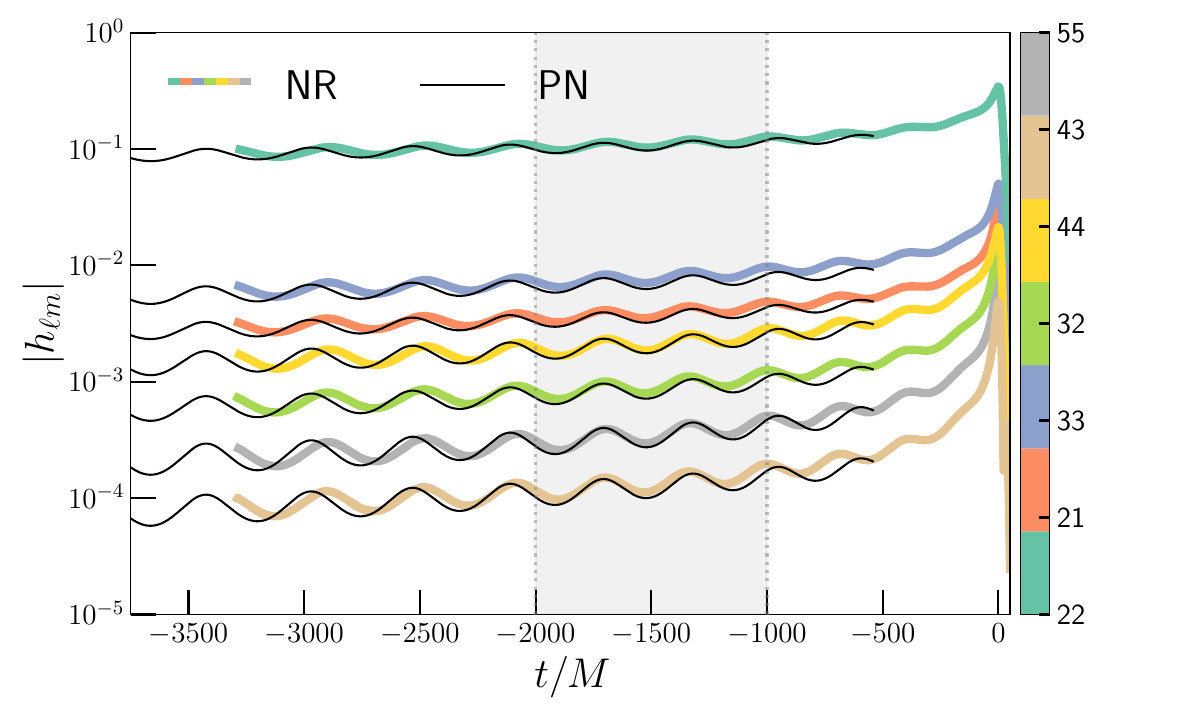}
    \includegraphics[width=0.49\textwidth]{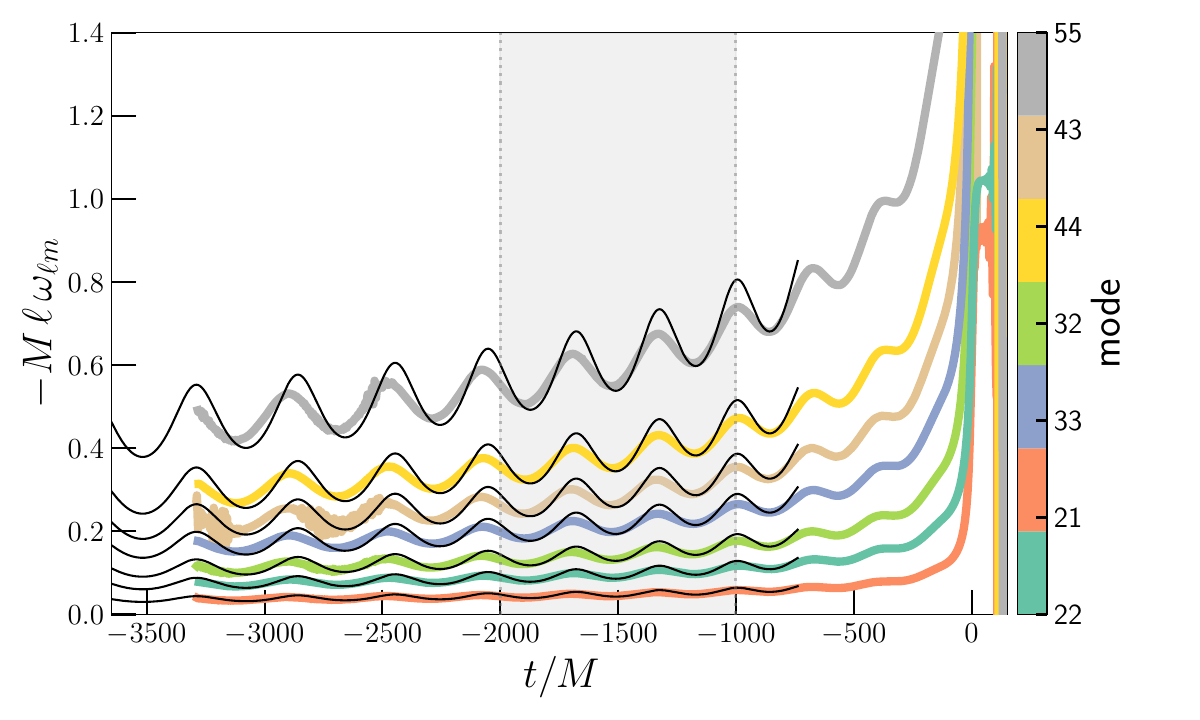}
    \caption{ 
    Amplitude and frequency of selected modes from an eccentric NR simulation (SXS:BBH:1364) together with an eccentric PN model are plotted. The PN model is evolved assuming a fixed value for initial eccentricity ($e_0=0.172$), mean anomaly ($l_0=2.681$) and the frequency-dependent PN parameter ($x_0=0.0391$).    
    The merger time of the NR waveform is set to zero. The initial set ($e_0, l_0, x_0$) is obtained by maximizing the overlap between the PN and NR waveform in a $1000M$ wide time window, in a region where the PN model is expected to agree with NR simulation. Additionally, a time shift is performed on the PN inspiral. The time window of maximum overlap is shown as the shaded region. The binary's component mass ratio ($q$) is 2, while the total mass ($M$) and the luminosity distance ($D_{\rm L}$) of the binary are set to $M$=1$M_{\odot}$ and $D_{\rm L}$=1Mpc respectively, following the convention of SXS simulations.
    }
    \label{fig:pn_nr_comparison}
\end{figure*}

Before we proceed to the technical part of the paper, we wish to highlight some of the important differences between the current work and Paper I. First, in Paper I we did not concern ourselves with possible differences between the eccentricity definitions and/or how they modify the waveforms. However, as argued above as well as in Refs.~\cite{Ramos-Buades:2022lgf, Shaikh:2023ypz}, in order to meaningfully compare two different models we must use a physically motivated definition of eccentricity. This, in fact is the primary motivation for our current work and modifies all input waveforms as well as the models constructed here.\footnote{Note also that, while in Paper I, $e_0$ was chosen in such a way that the model produces a value of $e_{\rm ref}$ at a given $x_{\rm ref}$ obtained in Ref.~\cite{Hinder:2017sxy}, the current work uses a gauge-independent eccentricity estimator (Ref.~\cite{Shaikh:2023ypz}) and ensures, at the same time, an agreement with the NR data.} Second, in Paper I, the criterion for choosing the window for hybridization was a bit \textit{ad hoc} and was based on a simple visual inspection. Here, instead we use the match (maximized over an initial time and phase) as a quantitative measure to identify the suitable hybridization window.\footnote{Note that the time in a NR simulation is measured from the start of the simulation in Paper I and from the merger in the current work and hybridization windows are different in the two works. In fact, the current work hybridizes the model closer to the merger and thus maximizes the PN inputs which helps in situations when NR simulations are not long.} 
Other elements that are new here are – an alternate, dominant mode model presented in Sec.~\ref{sec:TaylorT2 model} and its HM version which includes additional modes compared to the HM model of Paper I (see Sec.~\ref{sec:higher-mode-model} for details). The alternate model (referred to as \textsc{TaylorT2} model) is roughly 1.5 times faster than the model constructed in Sec.~\ref{sec:22-mode-model} (\textsc{EccentricTD} model) and also performs better (matches improve by roughly 2\% for eccentricities smaller than $\sim0.15$). The current work also addresses certain issues that remained unresolved for the model presented in Paper I. These are discussed in Sec.~\ref{sec:disc}.      
\\

The paper is structured in the following manner. In Sec.~\ref{sec:methods} we start by comparing waveforms from PN and NR approaches and discuss the construction of target hybrids. Next, in Sec.~\ref{sec:22-mode-model} we construct the waveform model by combining an eccentric PN inspiral model with a quasicircular merger-ringdown model at a suitable point obtained by performing comparisons with target models. Subsequently, the model is validated against the target models not used in calibrating it as well as against an independent family of waveforms. In Sec.~\ref{sec:alt_model}, we discuss an alternate model based on the prescriptions for attachment times obtained in Sec.~\ref{sec:22-mode-model} and subsequently extend this alternate model to include higher order modes. 
Finally, Sec.~\ref{sec:disc} presents summary of results and conclusions.

\section{Construction of target models with PN and NR inputs}
\label{sec:methods}

\begin{table}
\begin{tabular}{cllcccl}
  \hline
  \hline
  Count & Simulation ID & $q$ & $x_{0}$ & $e_0$ & $l_0$ & $N_{\rm orb}$\\
\hline
%1 & SXS:BBH:1132 & 1 & 20.56 & 0.000 & 2.852 & 53.3 & -\\
\multicolumn{7}{c}{\textbf{Training Set}}\\
1 & HYB:SXS:BBH:1355 & 1 & 0.0389 & 0.173 & 2.455 & 63.0\\
2 & HYB:SXS:BBH:1356 & 1 & 0.0375 & 0.230 & 1.717 & 65.5\\
3 & HYB:SXS:BBH:1358 & 1 & 0.0340 & 0.322 & 1.215 & 69.5\\
4 & HYB:SXS:BBH:1359 & 1 & 0.0347 & 0.317 & 1.131 & 67.0\\
5 & HYB:SXS:BBH:1360 & 1 & 0.0317 & 0.416 & 0.796 & 64.0\\
6 & HYB:SXS:BBH:1361 & 1 & 0.0313 & 0.416 & 0.796 & 66.0\\
7 & HYB:SXS:BBH:1364 & 2 & 0.0391 & 0.172 & 2.681 & 69.0\\
8 & HYB:SXS:BBH:1365 & 2 & 0.0376 & 0.209 & 2.262 & 72.5\\
9 & HYB:SXS:BBH:1366 & 2 & 0.0344 & 0.320 & 1.299 & 74.0\\
10 & HYB:SXS:BBH:1367 & 2 & 0.0346 & 0.320 & 1.299 & 73.5\\
11 & HYB:SXS:BBH:1368 & 2 & 0.0338 & 0.324 & 1.382 & 77.5\\
12 & HYB:SXS:BBH:1372 & 3 & 0.0344 & 0.300 & 1.789 & 90.0\\
13 & HYB:SXS:BBH:1373 & 3 & 0.0344 & 0.300 & 1.789 & 89.0\\
\multicolumn{7}{c}{\textbf{Testing Set}}\\
14 & HYB:SXS:BBH:1357 & 1 & 0.0344 & 0.322 & 1.215 & 67.5\\
15 & HYB:SXS:BBH:1362 & 1 & 0.0328 & 0.483 & 0.464 & 48.5\\
16 & HYB:SXS:BBH:1363 & 1 & 0.0308 & 0.505 & 0.590 & 51.5\\
%11 & HYB:SXS:BBH:1167 & 2 & 20.56 & 0.000 & 1.308 & 48.4 & -\\
17 & HYB:SXS:BBH:1369 & 2 & 0.0329 & 0.478 & 0.545 & 52.5\\
18 & HYB:SXS:BBH:1370 & 2 & 0.0291 & 0.508 & 0.628 & 63.0\\
%19 & HYB:SXS:BBH:1221 & 3 & 20.56 & 0.000 & 2.461 & 56.8 & -\\
19 & HYB:SXS:BBH:1371 & 3 & 0.0380 & 0.204 & 2.621 & 82.5\\
20 & HYB:SXS:BBH:1374 & 3 & 0.0290 & 0.495 & 0.832 & 77.5\\
\addlinespace[0.25cm]
\hline
\hline
% & HYB:SXS:BBH:0180 & 1 & 0.000 & 0.019 & 43.6\\
%1 & HYB:SXS:BBH:1355 & 1 & 0.120 & 1.423 & 41.7\\
%2 & HYB:SXS:BBH:1356 & 1 & 0.163 & 1.574 & 39.9\\
%3 & HYB:SXS:BBH:1357 & 1 & 0.227 & 0.451 & 36.1\\
%4 & HYB:SXS:BBH:1358 & 1 & 0.227 & -2.682 & 35.7\\
%5 & HYB:SXS:BBH:1359 & 1 & 0.227 & 1.834 & 36.3\\
%6 & HYB:SXS:BBH:1360 & 1 & 0.299 & -0.395 & 31.2\\
%7 & HYB:SXS:BBH:1361 & 1 & 0.299 & -1.019 & 31.0\\
%8 & HYB:SXS:BBH:1362 & 1 & 0.373 & -0.507 & 25.5\\
%9 & HYB:SXS:BBH:1363 & 1 & 0.373 & -0.912 & 25.3\\
%% & HYB:SXS:BBH:0184 & 2 & 0.000 & -2.821 & 48.5\\
%10 & HYB:SXS:BBH:1364 & 2 & 0.120 & -0.181 & 46.0\\
%11 & HYB:SXS:BBH:1365 & 2 & 0.145 & -1.127 & 44.9\\
%12 & HYB:SXS:BBH:1366 & 2 & 0.228 & -2.890 & 39.6\\
%13 & HYB:SXS:BBH:1367 & 2 & 0.228 & 1.687 & 40.5\\
%14 & HYB:SXS:BBH:1368 & 2 & 0.228 & 0.420 & 40.3\\
%15 & HYB:SXS:BBH:1369 & 2 & 0.373 & -0.203 & 28.3\\
%16 & HYB:SXS:BBH:1370 & 2 & 0.373 & 3.063 & 28.8\\
% & HYB:SXS:BBH:0183 & 3 & 0.000 & -3.121 & 56.7\\
%17 & HYB:SXS:BBH:1371 & 3 & 0.142 & 0.665 & 52.8\\
%18 & HYB:SXS:BBH:1372 & 3 & 0.209 & 3.005 & 48.7\\
%19 & HYB:SXS:BBH:1373 & 3 & 0.209 & 1.682 & 48.6\\
%20 & HYB:SXS:BBH:1374 & 3 & 0.359 & 3.114 & 35.0\\
%SXS:BBH:1355 & 1 & 0.120 & \hspace{3cm} SXS:BBH:1368 & 2 & 0.228 \\
%SXS:BBH:1356 & 1 & 0.163 & \hspace{3cm} SXS:BBH:1369 & 2 & 0.373\\
%SXS:BBH:1357 & 1 & 0.227 & \hspace{3cm} SXS:BBH:1371 & 3 & 0.142 \\
%SXS:BBH:1362 & 1 & 0.373 & \hspace{3cm} SXS:BBH:1373 & 3 & 0.209 \\
%SXS:BBH:1364 & 2 & 0.120 & \hspace{3cm} SXS:BBH:1374 & 3 & 0.359 \\  
\end{tabular}\vspace{-0.6cm}
\caption{
Set of time-domain hybrids constructed by matching NR simulations from the SXS catalog and state-of-the-art PN prescriptions for BBHs on eccentric orbits are listed. SXS simulation IDs are retained for identification with NR simulations used in constructing the hybrids. Each hybrid assumes a fixed value for initial eccentricity ($e_0$), mean anomaly ($l_0$) and the frequency-dependent PN parameter ($x_0$)  obtained using the {\fontfamily{qcr}\selectfont{GW\_ECCENTRICITY}} package based on Ref.~\cite{Shaikh:2023ypz}. The PN parameter ($x$) is related to the GW frequency ($f_{\rm gw}$) of the dominant mode as $x$ =$(\pi M f_{\rm gw})^{2/3}$ with $M$ representing the binary's total mass. Mass ratio ($q$) and number of orbits prior to the merger are also listed. $N_{\rm orb}$ is computed by taking the phase difference between the start of the waveform and the peak of the dominant mode ($\ell=2$, $|m|=2$) amplitude. 
}
\label{tab:nrsims}
\end{table}

\begin{figure*}[t!]
    \centering
    
    \includegraphics[width=0.45\textwidth]{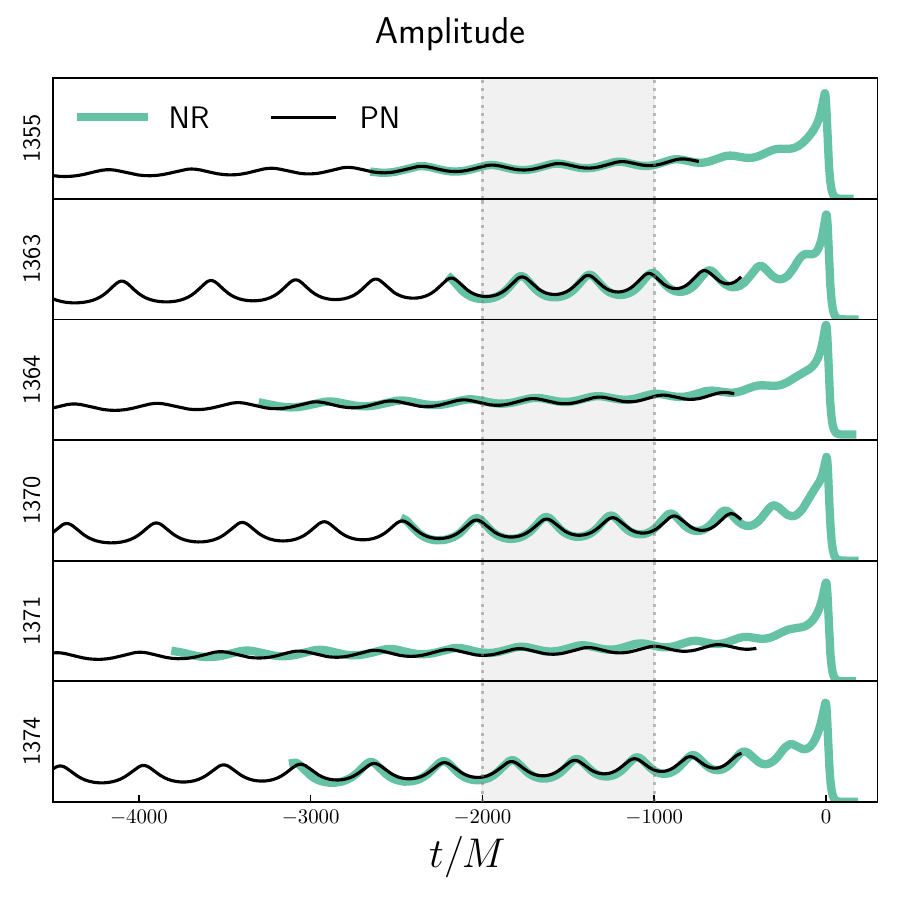}
    \includegraphics[width=0.45\textwidth]{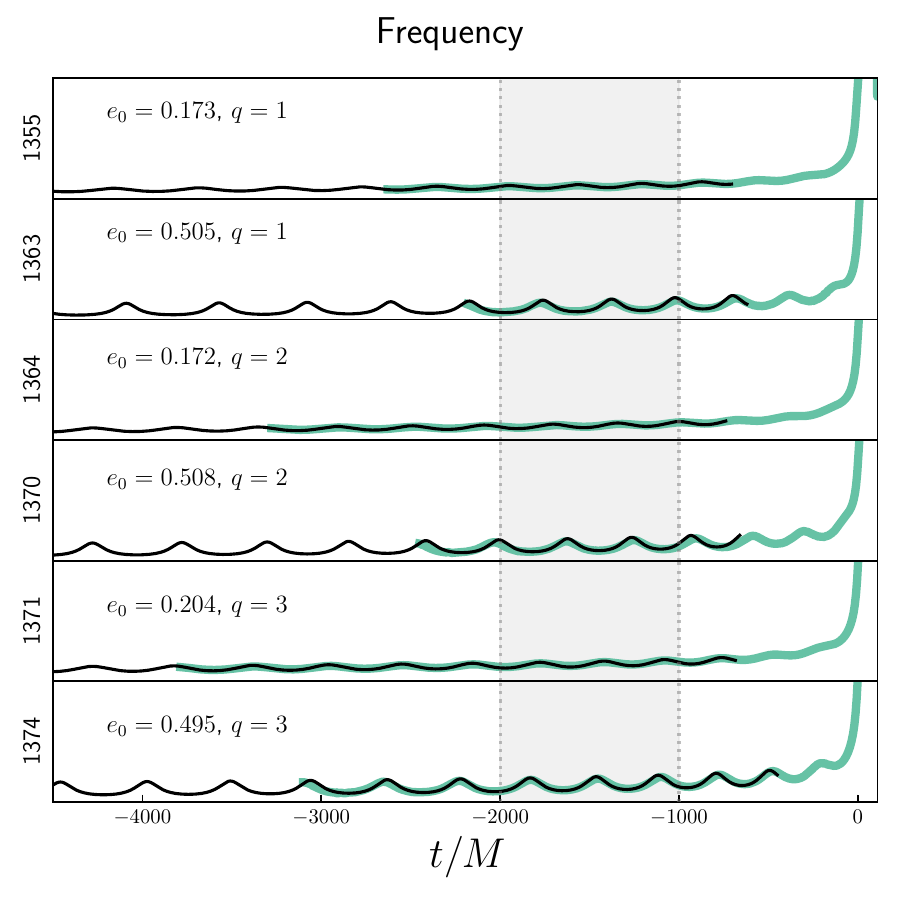}
    \caption{Same as Fig.~\ref{fig:pn_nr_comparison}, except that only dominant mode data are plotted and the comparison is shown for few other simulations with varying mass ratios and orbital parameters (see Table~\ref{tab:nrsims} for details). Initial eccentricity $(e_0)$ and mass ratio $(q)$ values for corresponding simulations are displayed in the right panel.
    Initial values of other parameters $(l_0, x_0)$ with which the PN model is evolved for comparisons with various simulations are listed in Table~\ref{tab:nrsims}.}
    
    \label{fig:22 mode comparison.}
\end{figure*}

\begin{figure*}[htbp!]
    \includegraphics[width=0.8\textwidth]{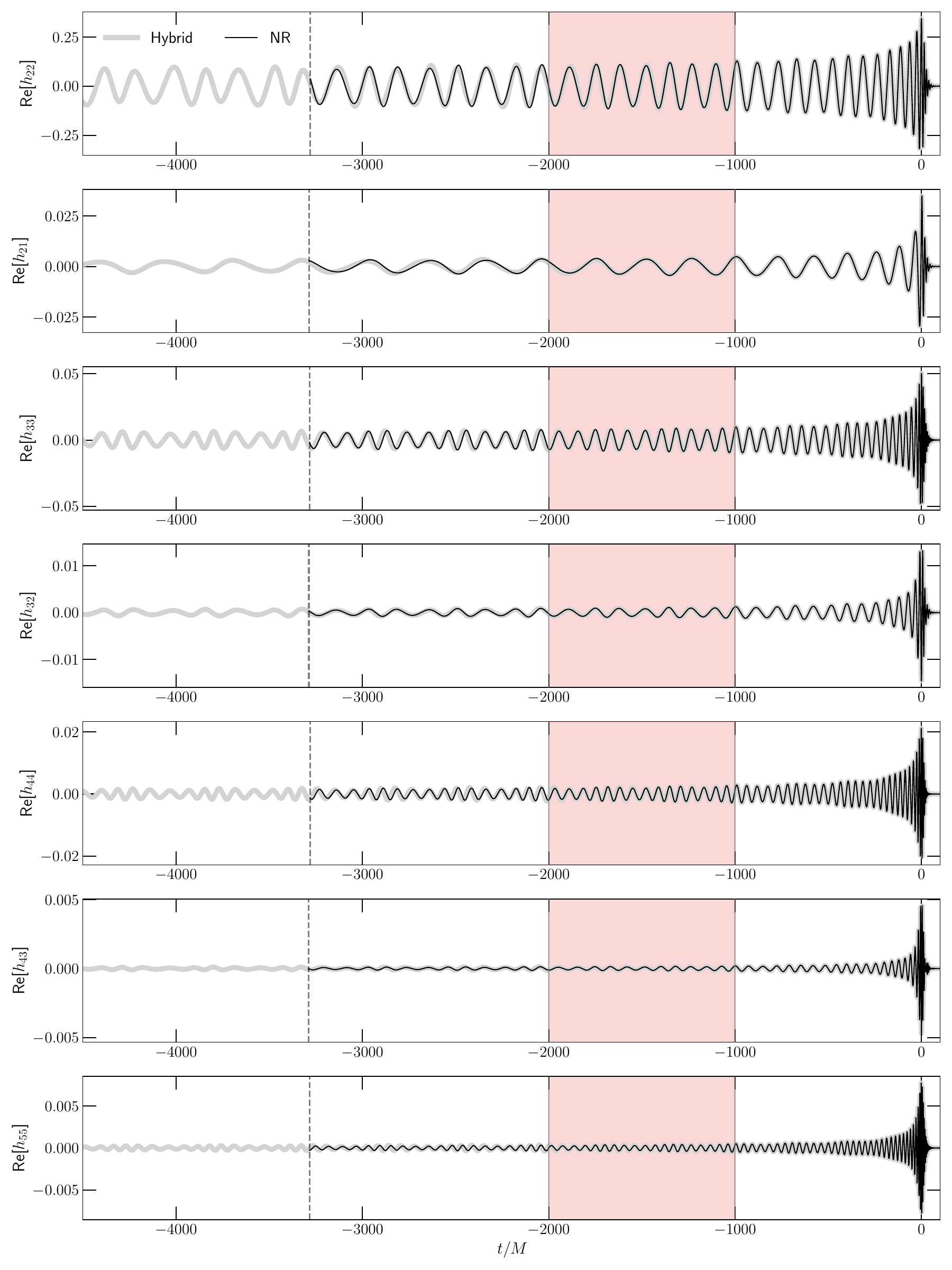}
    \caption{A hybrid model constructed by matching a NR simulation (SXS:BBH:1364) with a PN model (evolved using parameters consistent with the simulation) in a time window where the two are expected to correctly predict the binary dynamics. For comparison the NR data are also plotted. 
    The black dashed line marks the beginning of the NR waveform and the shaded light-red region $t \in (-2000M, -1000M)$ shows the matching window. Overlapping hybrid and NR waveforms on the left of the matching window hint at the quality of hybridization performed here. Table~\ref{tab:nrsims} lists details of all hybrids considered in the current work. 
    }
    \label{fig:hybridplot}
\end{figure*}
\subsection{PN and NR comparisons}
\label{sec:pn_inspiral_and_hybrids}
Paper I compares the PN inspiral waveforms with NR simulations. The inspiral mode amplitudes constituting 3PN inspiral waveforms [and eccentricity corrections up to \( O(e^6) \))] assuming nonspinning binary systems on quasielliptical orbits were computed in Refs.~\cite{Mishra:2015bqa,Boetzel:2019nfw,Ebersold:2019kdc}. The 2PN accurate orbital phase [up to \( O(e^6) \)] was taken from Ref.~\cite{Tanay:2016zog}.\footnote{The notation \( O(e^{6}) \) indicates that corrections in eccentricity are included up to the sixth power in eccentricity.}  
GW frequency for each PN mode was obtained using the following scaling relation \cite{Ramos-Buades:2022lgf}: 
\begin{equation}
    \omega_{\lm} \sim \frac{m}{2} \times \omega_{\rm 22}\ .
    \label{eq:omega_lm}
\end{equation}
The 20 eccentric, nonspinning NR simulations used for comparison with PN models were produced using the Spectral Einstein Code (SpEC) developed by  Simulating eXtreme Spacetimes (SXS) Collaboration and are publicly available \cite{Hinder:2017sxy, Boyle:2019kee}.
Comparison of PN and NR models for a specific simulation was shown in Fig. 1 of Paper I. Subsequently, a common region of validity was identified in which the two could be matched suitably to obtain hybrids listed in Table I there. (see Sec.s II A-II C of Paper I for technical details).\\ 

Here too we aim to compare the PN and NR prescriptions leading to the construction of hybrids. While the hybridization method is the same as in Paper I, comparison (which leads to identification of a suitable window for hybridization) is performed following a slightly different approach.
In Paper I, PN models were simply evolved to match a set of reference orbital parameters ($e_{\rm ref}, l_{\rm ref}$) computed at a reference GW frequency ($x_{\rm ref}$) computed in Ref.~\cite{Hinder:2017sxy} for each of the 20 simulations considered in Paper I. Here, we simply choose to compare the two waveforms in a $1000M$ wide time window, slide it over the overlapping data (from the start of the NR
waveform to the time corresponding to last-stable-orbit
frequency) and compute overlaps [the match maximized over a reference time and phase shifts; see Eq.~\eqref{match_def} below] by varying the input parameter trio ($e_{\rm ref},l_{\rm ref}, f_{\rm ref}$)
for the PN model, where $f_{\rm ref}$ is related to the PN parameter $x_{\rm ref}$ via $x_{\rm ref}$ =$(\pi M f_{\rm ref})^{2/3}$. 
The match (${\mathcal M}$) between two waveforms is defined as an inner product given as
\begin{equation}
    \mathcal{M} (\boldsymbol{\theta_1,\theta_2}) \equiv \mathrm{max}_{\phi_{\rm c},t_{\rm c}} \langle h(\boldsymbol{\theta_1}), h(\boldsymbol{\theta_2}) e^{i(2\pi ft_{\rm c} - \phi_{\rm c})} \rangle\,, 
    \label{match_def}
\end{equation}
\noindent with,
\begin{equation}
    \langle h_1,h_2 \rangle \equiv 4\ \mathrm{Re} \left[\int_0^\infty df \frac{\Tilde{h_1}^*(f) \Tilde{h_2}(f)}{S_{h}(f)}\right]\,,
\end{equation}
where $\langle h_1, h_2 \rangle$ represents the inner product between two waveforms $h_1$ and $h_2$ having unit norm and are functions of an intrinsic set of binary parameters ($\boldsymbol{\theta_1, \theta_2}$).
The phase $\phi_{\rm c}$ and time $t_{\rm c}$ are measured at coalescence and $S_{h}(f)$ represents the noise in the advanced LIGO detector~\cite{PhysRevD.102.062003, Owen:1998dk}.\\

We find the overlap is optimal for a time window of ($-2000M$ to $-1000M$) and for a given ($e_{\rm ref},l_{\rm ref}, f_{\rm ref}$) trio which becomes the starting reference ($e_0$, $l_0$, $f_0$) and are different for different NR simulations (see Table~\ref{tab:nrsims}).\footnote{We choose to work with a criterion of minimal match $\sim 96\%$ to identify whether or not the two prescriptions give consistent predictions about binary dynamics.} Compared to the earlier approach, our current approach helps the construction of hybrids in at least two distinct ways. (1) The identification of hybridization window is done using quantitative measures such as overlaps and (2) the reference orbital parameters such as orbital eccentricity ($e_0$) and mean anomaly ($l_0$) 
at a given frequency ($f_0$) will be free from gauge ambiguities due to the use of the definition of Refs.~\cite{Shaikh:2023ypz, Ramos-Buades:2022lgf}. With the suitable hybridization window identified, we can now proceed to reconstruct the hybrids.\\

Figure~\ref{fig:pn_nr_comparison} compares the data corresponding to the NR simulation bearing simulation ID SXS:BBH:1364, for a selected set of modes chosen based
on their relative significance compared to the dominant
mode.\footnote{See a discussion in Sec.~II B of Paper I for specific details on how these modes are identified and we simply stick to the choice there.} 
The PN model is evolved using a set of initial parameters ($e_0$, $l_0$, $f_0$) obtained using the procedure discussed above and then plotted together with the NR simulation after performing a time shift. The window giving maximum match is also displayed. For completeness we also show similar comparisons for few other simulations in Fig.~\ref{fig:22 mode comparison.} albeit for only the dominant mode. Interestingly, we observe that the time window returning maximum match is common for all simulations.

\subsection{Construction of hybrid waveforms}
\label{sec:hybrid_waveforms}

Complete IMR waveforms are constructed by matching PN and NR prescriptions for a set of modes included in Fig.~\ref{fig:pn_nr_comparison}, in a region where the PN prescription closely mimics the NR data following the method of Ref.~\cite{Varma:2016dnf}. These are traditionally referred to as ``hybrids". As discussed in Ref.~\cite{Varma:2016dnf}, construction of hybrids including higher modes (in the circular case) is possible by performing at least two rotations (and a time shift) so as to align the frames in which PN/NR waveforms are defined.\footnote{The third Euler angle is assumed to be oriented in the direction of the total angular momentum of the binary
(see Fig.~2 and the discussions in Sec. III C of Ref.~\cite{Varma:2014jxa}).  Note that, footnote 5 of Paper I uses the incorrect reference. The correct reference is Ref.~\cite{Varma:2014jxa}.
} This argument was simply extended to the case of eccentric orbits in Paper I, assuming that the effect of marginalizing over parameters such as eccentricity and mean anomaly will not significantly affect the hybridization. As discussed above, for the current work we simply adopt the hybridization procedure of Paper I.\footnote{The hybridization technique can be modified using the prescriptions from Ref. \cite{Varma:2018mmi}, although it does not lead to significant changes in the waveforms.}
The prescription for construction of hybrids is discussed in detail in Ref.~\cite{Varma:2016dnf} as well as in Paper I. Nevertheless, we reproduce some of the steps here for completeness.\\ 

A least-squares minimization of the integrated difference between the GW modes from the PN and NR waveforms in a time interval ($t_{\rm i}, t_{\rm f}$), in which the two approaches give similar results, is performed and can be defined as  
\begin{equation}
\delta = \mathrm{min}_{t_0,\varphi_0, \psi} \int_{t_{\rm i}}^{t_{\rm f}} dt \sum_{\ell,m} \left|\hlm^\mathrm{NR}(t-t_0)e^{i(m\varphi_0+\psi)}-\hlm^\mathrm{PN}(t)\right|,
\label{eq:delta}
\end{equation}
where the minimization is performed over a time shift ($t_0$) and the two angles ($\varphi_0, \psi$) as discussed above. The hybrid waveforms are then constructed by combining the NR data with the ``best matched'' PN waveform in the following way:
\begin{equation}
\hlm^\mathrm{hyb}(t) \equiv \, \tau(t) \, \hlm^\mathrm{NR}(t-t_0') \ e^{i(m\varphi_0'+\psi')} + (1-\tau(t)) \, \hlm^\mathrm{PN}(t) ,
\label{eq:hyb}
\end{equation}
where ($t_0'$, $\varphi_0'$, $\psi'$) are the values of ($t_0$, $\varphi_0$, $\psi$) that minimize the integral of Eq.~\eqref{eq:delta}. In the above equation, $\tau(t)$ is a weighting function defined by
\begin{eqnarray}
\tau(t) \equiv \left\{ \begin{array}{ll}
0 & \textrm{if $t < t_{\rm i} $}\\
\frac{t-t_{\rm {i}}}{t_{\rm {f}}-t_{\rm {i}}}  & \textrm{if $t_{\rm{i}} \leq t < t_{\rm{f}} $}\\
1 & \textrm{if $t_{\rm{f}} \leq t$.}
\end{array} \right.
\label{eq:tau}
\end{eqnarray}

The hybrids corresponding to a representative NR simulation (SXS:BBH:1364) for all relevant modes are shown in Fig.~\ref{fig:hybridplot}.
The two waveforms are aligned at merger and the shaded gray region $t \in (-2000M, -1000M)$ highlights the matching window where hybridization was performed. Overlapping hybrid and NR waveforms outside (on the left of) the matching window hint at the quality of hybridization performed here.\\ 

We reconstruct IMR hybrids corresponding to all $20$ eccentric NR simulations listed in Ref. \cite{Hinder:2017sxy} as was done for Paper I. These are listed in Table~\ref{tab:nrsims} and the SXS simulation IDs have been retained to identify the hybrids with the corresponding NR simulation. Each hybrid starts with a specific initial eccentricity $(e_0)$, mean anomaly ($l_0$) and frequency $(x_0)$ obtained following the procedure discussed in Sec.~\ref{sec:pn_inspiral_and_hybrids}.

\begin{figure*}[ht!]
   \centering
       \includegraphics[width=0.9\textwidth]{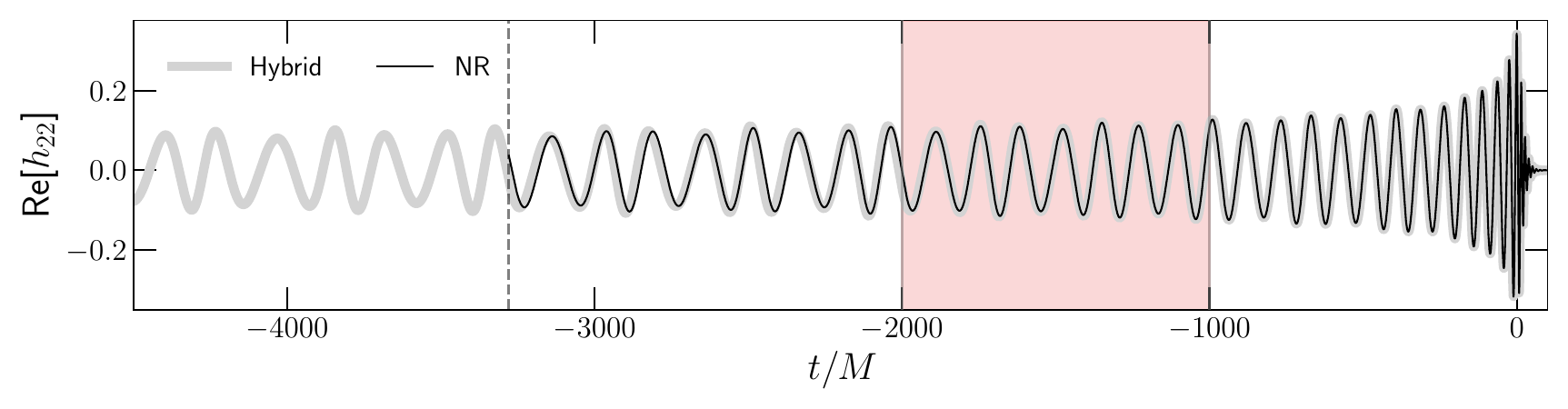}
   \caption{Same as Fig.~\ref{fig:hybridplot}, except the PN part of the hybrid is purely based on \textsc{EccentricTD} model of Ref.~\cite{Tanay:2016zog}. These are also the hybrids that are used in constructing the model in Sec.~\ref{sec:22-mode-model}. 
   } 
   \label{fig:NR-hyb-EccTD}
\end{figure*}

\section{The waveform model}
\label{sec:22-mode-model}

\begin{figure*}[ht!]
    %\centering
    \includegraphics[height=4cm, width=0.48\textwidth,keepaspectratio ]{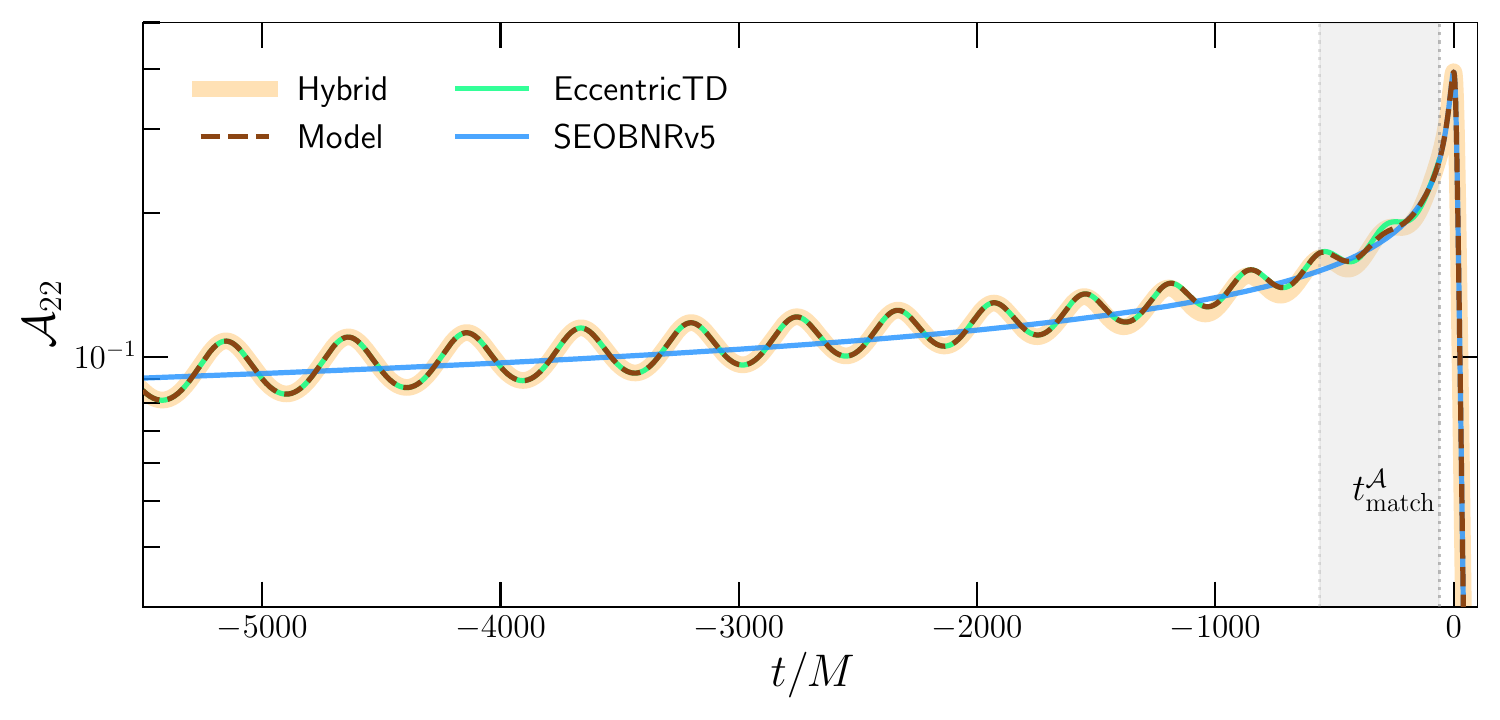}
    \includegraphics[height=4cm,width=0.48\textwidth, keepaspectratio]{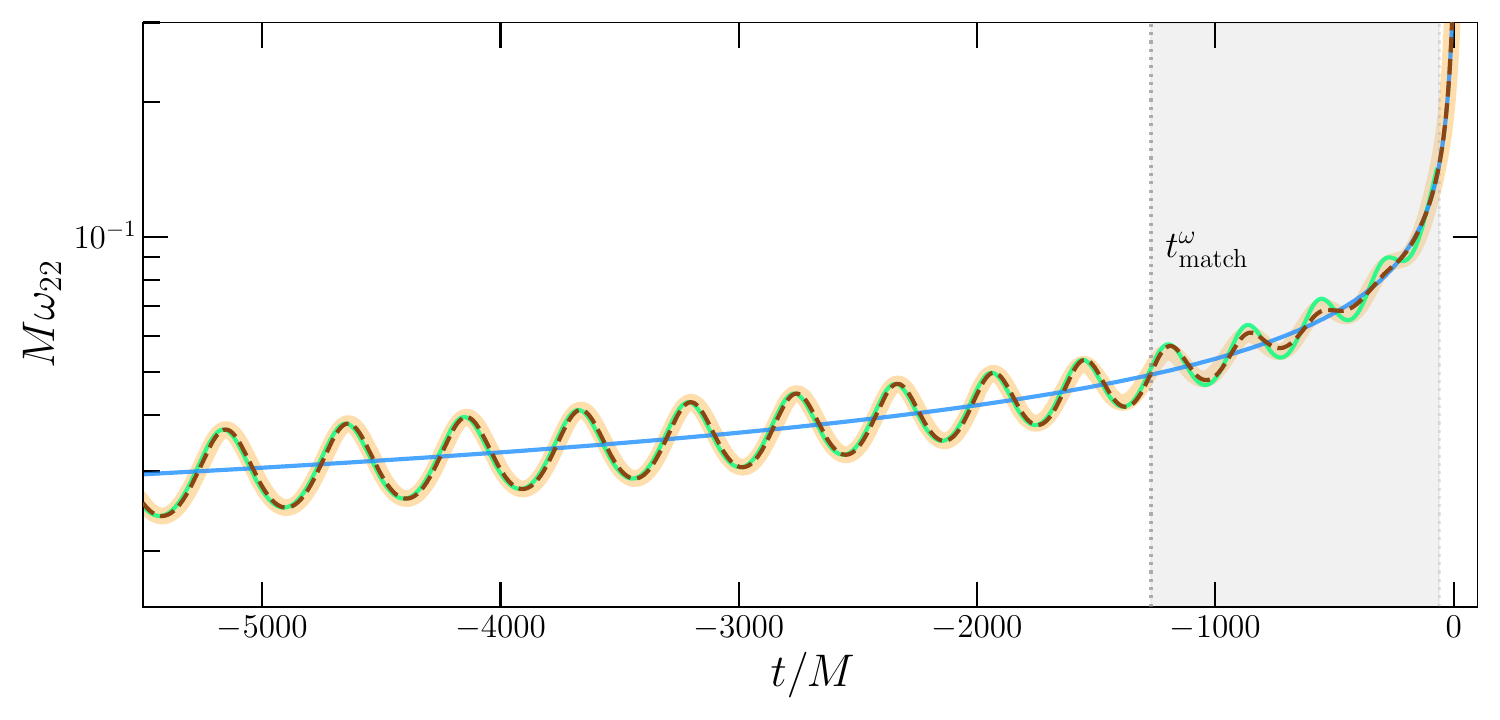}
    \caption{The (numerical) amplitude and frequency model  produced by combining an eccentric inspiral with a quasicircular merger-ringdown waveform for a total mass of 30$M_{\odot}$ is plotted against the hybrid (developed in Sec.~\ref{sec:22-mode-model}) used in calibrating the model. The eccentric inspiral (\textsc{EccentricTD}~\cite{Tanay:2016zog}) and the (quasicircular) merger-ringdown (\textsc{SEOBNRv5}~\cite{Pompili:2023tna}) models are also plotted for comparison. The amplitude (frequency) model transitions smoothly from the inspiral to the merger-ringdown stage inside the shaded region(s), at $t_{\rm match}^\mathcal{A}$ ($t_{\rm match}^\omega$) values, maximizing the overlap between the model and the target hybrid.
    }
    \label{fig:td-model-amp}
\end{figure*}
Until now we focused on constructing a (PN-NR) hybrid model which could be used as a target for building a fully analytical IMR model for eccentric binary black hole mergers.\footnote{Note that these hybrids are simply the longer versions of NR simulations with fixed component mass ratios and are only scalable by the binary's total mass and distance from the observer.}
While these hybrids could be used to construct the model following the procedure adopted in Paper I, as was done there, we construct an independent set of hybrids using the PN model, \textsc{EccentricTD}~\cite{Tanay:2016zog} (see Fig.~\ref{fig:NR-hyb-EccTD}). This is primarily done to minimize the difference between the target hybrids and the model (being constructed) that also uses the waveform \textsc{EccentricTD}~\cite{Tanay:2016zog} for the inspiral part.\footnote{The inspiral part of the model (\textsc{EccentricTD}~\cite{Tanay:2016zog}) is evolved using $e_0$ at a given $x_0$, taken from Table~\ref{tab:nrsims}, as inputs. The motivation here is to use only one set of (physically motivated) values for $e_0$ to evolve models with varying parametrizations at a small cost to the accuracy of resulting waveforms. In fact, we find the largest change in the $e_0$ estimates is of the order of 1\% which does not impact the overall performance of the model.} Note that, these new hybrids only include the dominant mode ($\ell=2, |m|=2$) since it is the dominant mode that we wish to model first. Note that, we do not use these hybrids in validating the model (See Fig.~\ref{fig:match_ecc_temp_new}) constructed in this section.\\

As in Paper I, here too we obtain a fully analytical dominant ($\ell=2, |m|=2$) mode model by matching an eccentric PN inspiral \cite{Tanay:2016zog} with a quasi circular prescription for the merger-ringdown phase \cite{Pompili:2023tna}.
Here too we stick to the procedures adopted in Paper I which involve identifying attachment times for both amplitude and frequency data together with an overall shift and the output is a coherent IMR model suitable for generating desired signals. Note however, that one must map the data for attachment times and time shifts to a set of physical parameters of the binary such as mass ratio and other relevant parameters. Exact details concerning this model are outlined in Secs.~\ref{sec:time-shift}-\ref{sec:frequency model}. For $M \sim 25M_\odot$, the highest $x_0$ value of Table~\ref{tab:nrsims} corresponds to a frequency of $f\sim20$Hz (low frequency cutoff for advanced LIGO design \cite{TheLIGOScientific:2014jea}). This motivates us to work with a conservative choice of a $30M_\odot$ system when constructing the model. We find only 13 out of 20 cases reasonably agree (with overlaps $>$ 97\%) for a $30M_\odot$ system with the hybrids and thus the final fits are obtained only using these 13 performing cases. Note however, that the PN inspiral is analytic and thus the model can be generated for arbitrary low masses.
\subsection{Numerical model}
\label{sec:numerical_model}
\subsubsection{Time shift}
\label{sec:time-shift}
The process of generating a numerical model is similar to the procedure adopted in Paper I and we reproduce it here for clarity and completeness. As described in Sec.~\ref{sec:hybrid_waveforms}, hybridization involves a minimization over a time shift, so when producing the amplitude model, we have to first perform a time shift of the inspiral waveform relative to the circular IMR waveform, because the time to merger is not known. This is done by first setting the merger time for the circular IMR waveform to zero and then time sliding the eccentric inspiral about the merger. We start by making a trial choice of $t_{\rm{shift}}$ and then generate an amplitude and a phase model by the methods described in Secs. \ref{sec:amplitude model} and \ref{sec:frequency model}, respectively.

\subsubsection{Amplitude model}
\label{sec:amplitude model}

As can be seen in Fig.~\ref{fig:pn_nr_comparison}, the waveforms tend to circularize near merger.\footnote{See also the discussion around Fig. 3 of Ref.~\cite{Hinder:2017sxy} which clearly shows all NR simulations become circular $30M$ before the merger.} Hence, in order to model this effect, we can join the eccentric inspiral to the circular IMR by  suitable choice of an appropriate time $t_{\rm{match}}^\mathcal{A}$. The amplitude model is obtained by joining the eccentric inspiral with the circular IMR using a transition function over a time window of $500M$ which ends at $t_{\rm{match}}^\mathcal{A}$. Given a target hybrid, and a trial choice of $t_{\rm{shift}}$, we start with a trial choice of $t_{\rm{match}}^\mathcal{A}$ roughly $500 M$ before the merger and produce the amplitude model as given below,

\begin{equation}
\mathcal{A}_{22}^\mathrm{model}(t) \equiv \, \tau_{\rm{a}}(t) \, \mathcal{A}_{22}^\mathrm{IMR}(t) \  + (1-\tau_{\rm{a}}(t)) \, \mathcal{A}_{22}^\mathrm{inspiral}(t) ,
\label{eq:amp_model}
\end{equation}
where $\tau_{\rm{a}}(t)$ is defined as 

\begin{eqnarray}
\tau_{\rm{a}}(t) \equiv \left\{ \begin{array}{ll}
0 & \textrm{if $t < t_{\rm{i}} $}\\
\frac{t-t_{\rm{i}}}{t_{\rm{f}}-t_{\rm{i}}}  & \textrm{if $t_{\rm{i}} \leq t < t_{\rm{f}} $}\\
1 & \textrm{if $t_{\rm{f}} \leq t$.}
\end{array} \right.
\label{eq:tau_amp}
\end{eqnarray}
We set $t_{\rm{i}} = t_{\rm{match}}^\mathcal{A} - 500M$ and $t_{\rm{f}} = t_{\rm{match}}^\mathcal{A}$ as the bounds of the time interval over which the two waveforms are joined. Figure~\ref{fig:td-model-amp} demonstrates the process. The gray region is the time interval ending at $t_{\rm{match}}^\mathcal{A}$ where the inspiral and circular IMR is joined.\\
\\
After the amplitude model is obtained for a particular choice of trial $t_{\rm{shift}}$ and $t_{\rm{match}}^\mathcal{A}$, we combine it with the target hybrid phase to obtain the polarizations and then calculate the match with the target hybrid.\footnote{This is done to ensure that the only component that is different between the target hybrid and the template model is the amplitude that we model here.} 
We then change the trial choice of $t_{\rm{match}}^\mathcal{A}$ by $5M$, bringing it closer to the merger, and repeat the process of producing the amplitude model, and calculating the match. This variation of $t_{\rm{match}}^\mathcal{A}$ is done until roughly $30M$ before merger. We thus obtain a set of match values for varying $t_{\rm{match}}^\mathcal{A}$ but for a single trial $t_{\rm{shift}}$ and pick the one that has the highest value of match. We repeat the exercise for other choices of $t_{\rm{shift}}$ (trial $t_{\rm{shift}}$ varies between $-400 M$ and $400 M$ in steps of $5 M$) and find out the corresponding $t_{\rm{match}}^\mathcal{A}$ with the highest value of match. Thus, we obtain a set of $t_{\rm{shift}}$ and  $t_{\rm{match}}^\mathcal{A}$ pairs with a match value for each pair. From this set, the pair with the highest value of match is chosen as the numerical estimate for $t_{\rm{shift}}$ and $t_{\rm{match}}^\mathcal{A}$ for a particular target hybrid. We obtain numerical estimates using the same process for all 20 target hybrids.

\subsubsection{Frequency model}
\label{sec:frequency model}

\begin{figure*}[htbp!]
   \centering
    \includegraphics[width=0.32\textwidth]{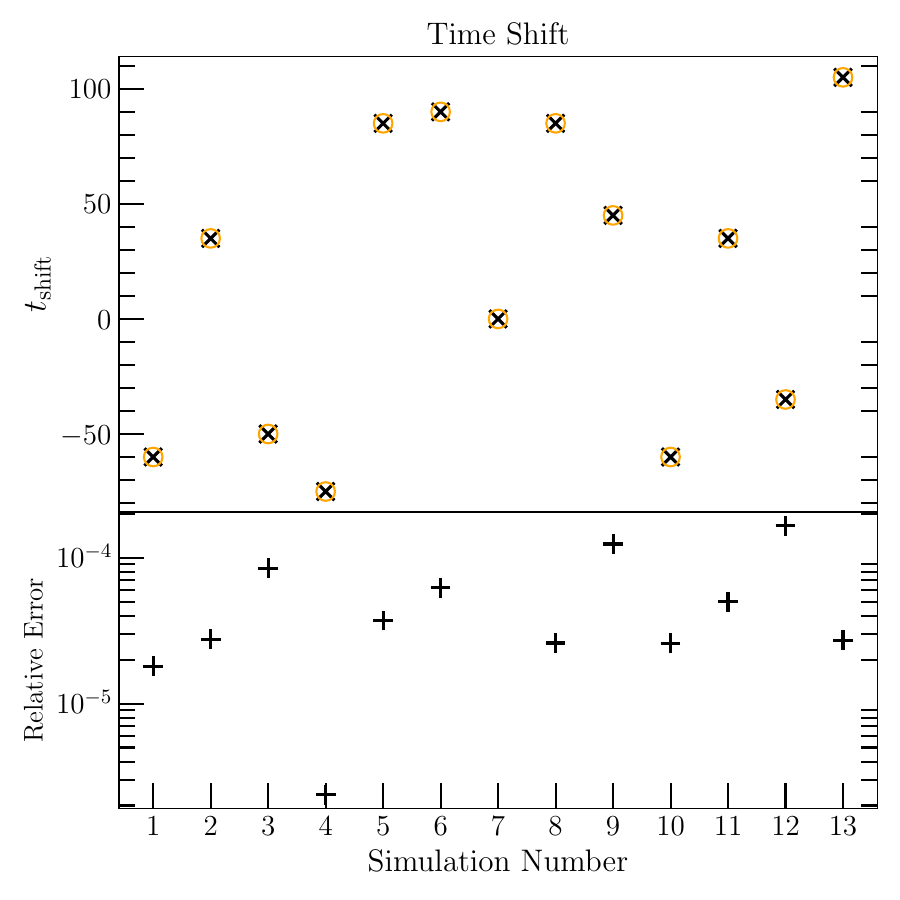}
    \includegraphics[width=0.32\textwidth]{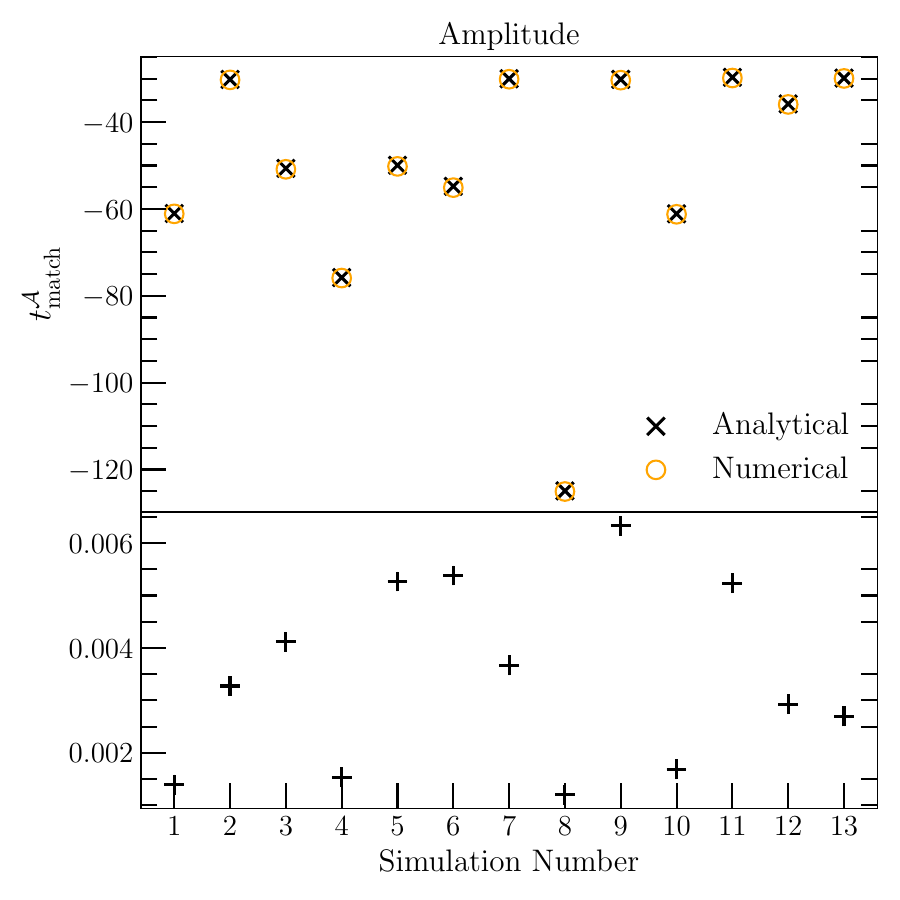}
    \includegraphics[width=0.32\textwidth]{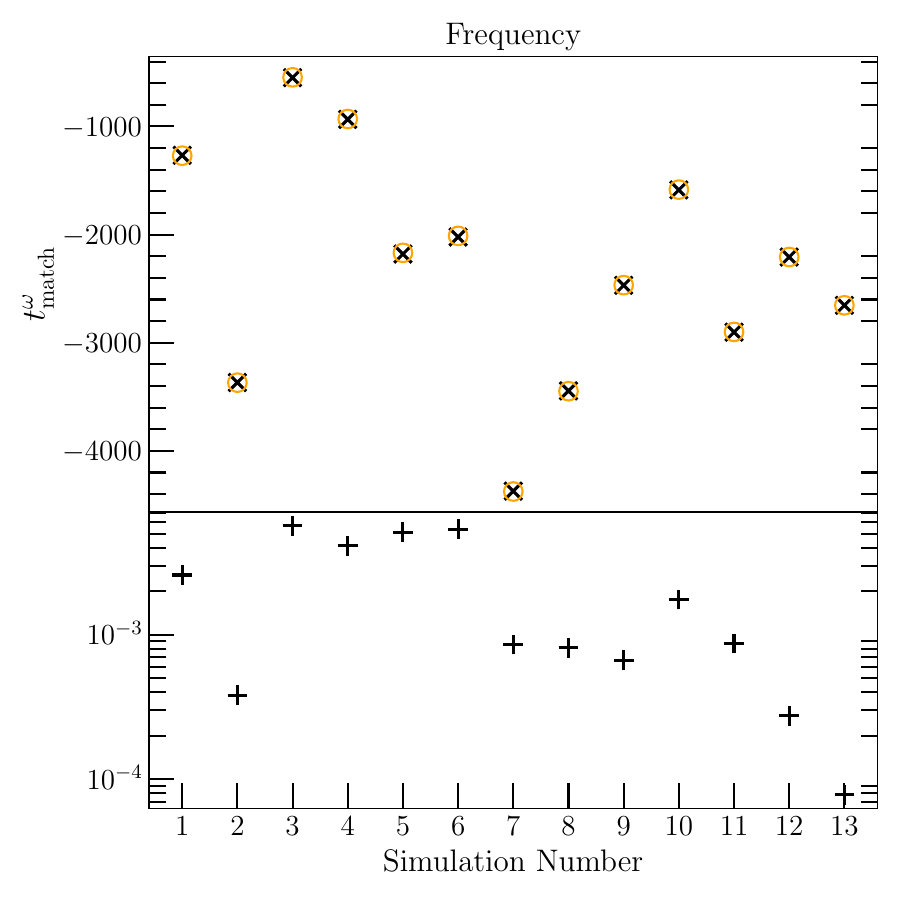}
   \caption{Top: numerical fits for $t_{\rm{match}}^\mathcal{A}$ and $t_{\rm{match}}^\omega$ as well as for $t_{\rm{shift}}$ are mapped into the physical parameter space for eccentric, nonspinning systems characterized by the binary's eccentricity, mean anomaly at a reference frequency and the mass ratio parameter $q$ or $\eta$ depending upon the model best fitting the data. Circles represent the numerical data points while crosses represent the value returned by the analytical best-fit model. Bottom: relative error between the numerical and analytical estimates are displayed.
   Simulation numbers on the x axis represent 13 training set hybrids listed in Table~\ref{tab:nrsims}. The best-fit model(s) predict the numerical estimates for attachment times and the time shift within $\pm 1M$.}
    \label{fig:amplitude_fit}
\end{figure*}

For the frequency model, we follow a similar procedure as described in Sec.~\ref{sec:amplitude model} with the only difference being the duration of the time interval where the inspiral frequency is joined with the circular IMR frequency. The value of $t_{\rm{shift}}$ is fixed to the one that was obtained while producing the amplitude model. Similar to the amplitude model procedure, we determine an appropriate $t_{\rm{match}}^\omega$ for joining the inspiral frequency with the circular IMR frequency. However, the time interval where the two are joined, starts at $t_{\rm{match}}^\omega$ and ends at a time close to $30M$ before merger.\footnote{This choice is motivated by the fact that all NR simulations considered in this work necessarily circularize $30M$ before the merger \cite{Hinder:2017sxy}.} Just like the amplitude model, we start with the choice of a trial value of frequency $t_{\rm{match}}^\omega$ roughly $6000M$ before merger and obtain the frequency model as given below,

\begin{equation}
\omega_{22}^\mathrm{model}(t) \equiv \, \tau_{\rm{a}}(t) \, \omega_{22}^\mathrm{IMR}(t) \  + (1-\tau_{\rm{a}}(t)) \, \omega_{22}^\mathrm{inspiral}(t) ,
\label{eq:frequency_model}
\end{equation}
where $\tau_{\rm{a}}(t)$ is as defined in Eq.~\eqref{eq:tau_amp} with the difference being $t_{\rm{i}} = t_{\rm{match}}^\omega$ and $t_{\rm{f}} \lesssim -30M$. Figure~\ref{fig:td-model-amp} demonstrates the process.\\
\\
Once the frequency model is obtained for the choice of trial $t_{\rm{match}}^\omega$, we calculate the phase by integrating the frequency model. This is then combined with the amplitude model obtained for the same target hybrid (generated using the numerical estimate of $t_{\rm{shift}}$ and $t_{\rm{match}}^\mathcal{A}$ already obtained) to produce the polarizations and a match with the target hybrid is calculated. We then change the trial choice of $t_{\rm{match}}^\omega$ by $1M$, bringing it closer to the merger and repeat the process of producing the frequency model, and calculating the match.\footnote{We use finer step size 
to vary the trial choice when searching for $t_{\rm match}^\omega$ (as opposed to $t_{\rm match}^\mathcal{A}$) as the match between the target hybrid and the model is more sensitive to a change in the $t_{\rm match}^\omega$ value (compared to $t_{\rm match}^\mathcal{A}$ value).
}
Once again, we do this variation until roughly $30M$ before merger to obtain a set of match values for varying $t_{\rm{match}}^\omega$ and pick the one that has the highest value of match. The corresponding value of frequency $t_{\rm{match}}^\omega$ is the numerical estimate for a particular target hybrid. We obtain numerical estimates for all 20 target hybrids using the same process.

\subsection{Analytical model}
\label{sec:analytical model}

\begin{figure*}[htbp!]
    \centering
    \includegraphics[width=0.49\textwidth]{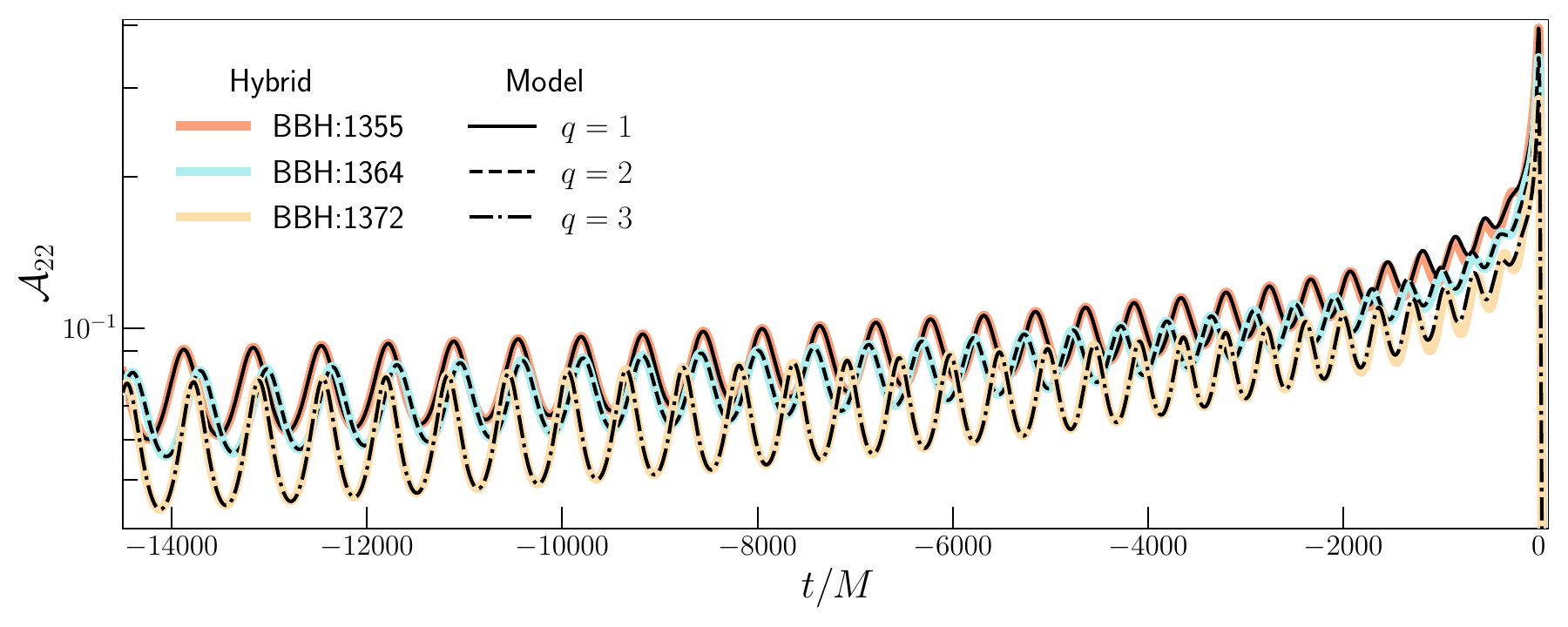}
    \includegraphics[width=0.49\textwidth]{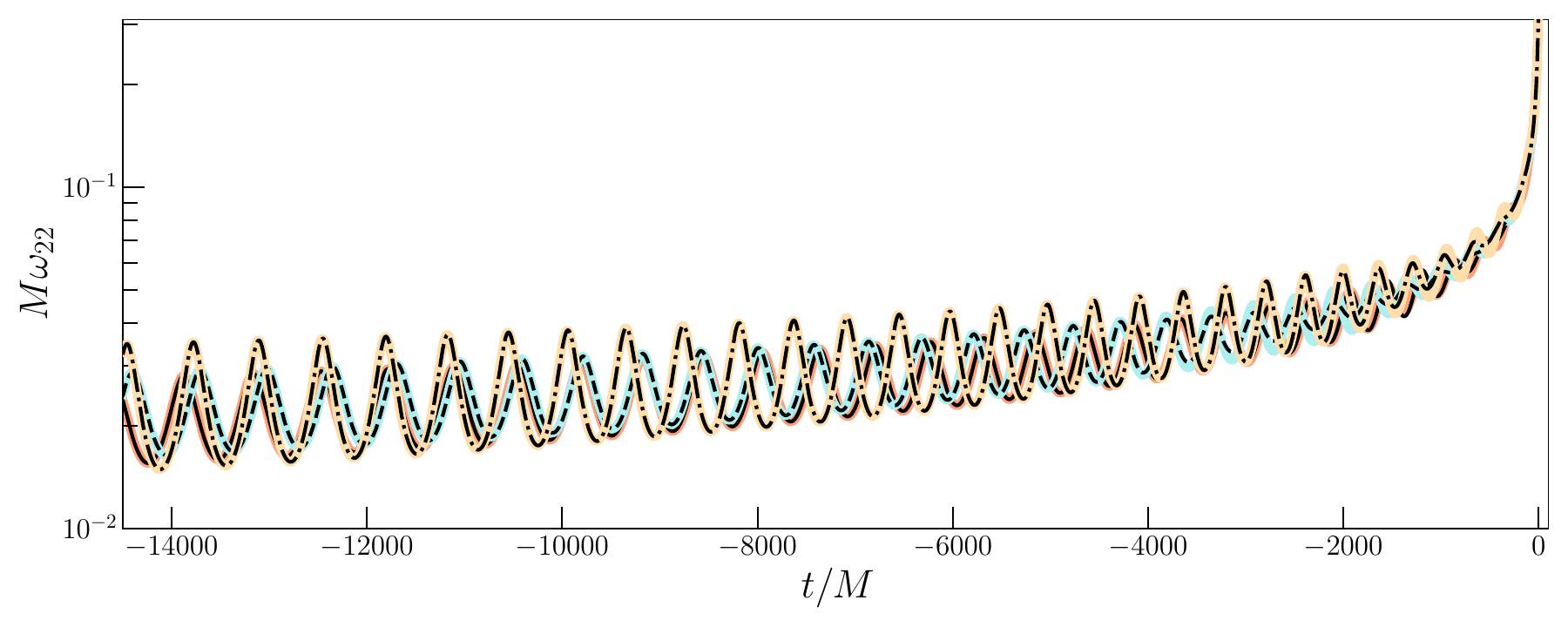}
    \includegraphics[height=4cm, width=0.98\textwidth, keepaspectratio]{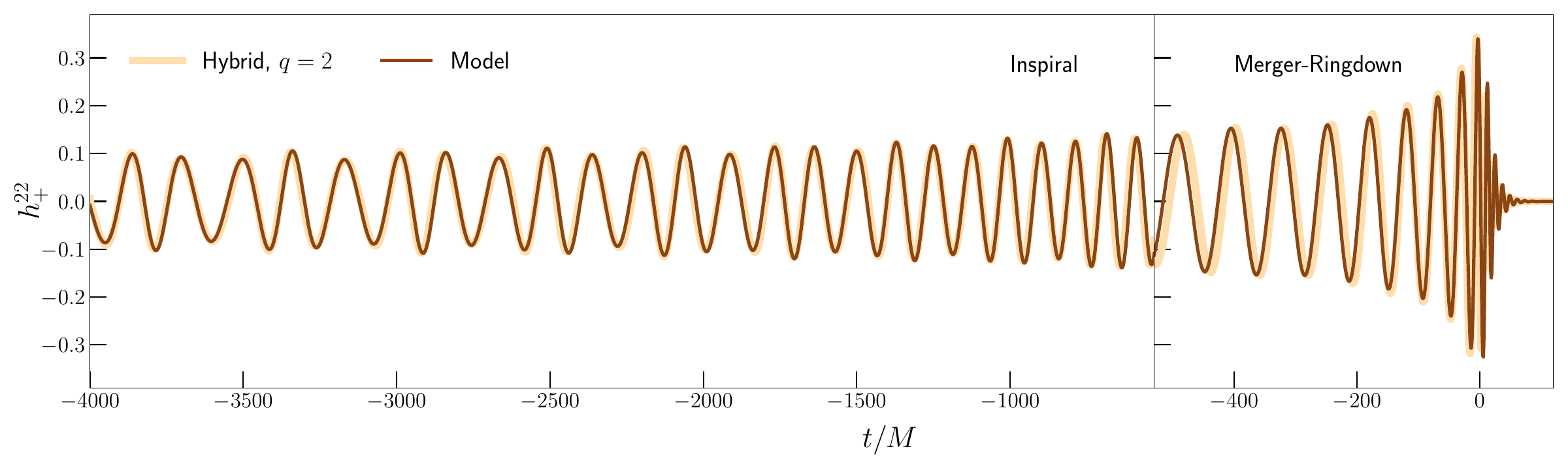}
    \caption{Top: dominant mode ($\ell$=2, $|m|$=2) amplitude and frequency models 
    are plotted against three representative target hybrids (developed in Sec.~\ref{sec:22-mode-model}) used in training the model. Bottom: one of the polarizations, obtained by combining the amplitude and the frequency model shown in the top panel for the $q=2$ case, is shown as a visual proof of the quality of the model being presented. The transition from inspiral to merger-ringdown is shown (for visual clarity) at $t_{\rm match}^\mathcal{A}$ given by Eq.~\eqref{eq:t_match_amplitude} and has no impact on the plotted data.
    }
    \label{fig:td-model-q123-amp}
\end{figure*}

We have described the procedure of producing (numerical) time-domain model fits for the dominant mode model, where we used a set of 20 eccentric hybrids as targets to calibrate our model. For each hybrid, we obtained a numerical estimate for $t_{\rm{shift}}$, $t_{\rm{match}}^\mathcal{A}$ and $t_{\rm{match}}^\omega$. In order to be able to generate waveforms for an arbitrary configuration these numerical fits need to be mapped into the physical parameter space for eccentric systems characterized by the binary's eccentricity, mean anomaly at a reference frequency and the mass ratio parameter. In this section, we determine a functional form by performing analytical fits to these numerical estimates. For analytical fits, we consider only those simulations (13 of the 20) for which the match between numerical model and the corresponding eccentric hybrid is greater than $97\%$ %{\red for a $30M_\odot$}
and collectively refer to them as the \textit{``training set"} and the remaining seven simulations are categorized as \textit{``testing set"} although only two of these (HYB:SXS:BBH:1357, 1371) can really be used to test the model as other simulations have initial eccentricities significantly larger than any of the training set hybrids and thus outside the calibration range for the model (see for instance, Table~\ref{tab:nrsims}). For this reason, when validating the model against hybrids we only include these two hybrids from the testing set. The fitted functions obtained are of the form as mentioned below, 

\begin{align}
t_{\rm{shift}}\,(q,e,l) &= \sum_{\alpha, \beta, \gamma, \delta} A_{\alpha \beta \gamma \delta} \, e^\alpha \, q^\beta \, \cos( \gamma \, l \,+\, \delta \,e \, l \,+\, a_{\alpha \beta \gamma \delta})\,,
\label{eq:t_shift_analytical}
\end{align}
for time shift, where $A_{\alpha \beta \gamma \delta} = a_{\alpha \beta \gamma \delta} = 0$ for $\alpha + \beta > 3$ and/or $\alpha > 2$ and/or $\gamma + \delta > 1$, and $A_{\alpha 0 0 1} = A_{0 \beta 1 0} = A_{0 \beta 0 1} = A_{0 0 \gamma \delta} = a_{\alpha \beta 0 0 } = 0$,

\begin{align}
t_{\rm{match}}^{\mathcal{A}} \,(\eta,e,l) &= \sum_{\alpha, \beta, \gamma, \delta} B_{\alpha \beta \gamma \delta} \, e^\alpha \, \eta^\beta \, \cos( \gamma \, l \,+\, \delta \,e \, l \,+\, b_{\alpha \beta \gamma \delta})\,,
\label{eq:t_match_amplitude}
\end{align}
for amplitude, where $\eta=q/(1+q)^2$,     $B_{\alpha \beta \gamma \delta} = b_{\alpha \beta \gamma \delta} = 0$ for $\alpha + \beta > 3$ and/or $\alpha > 2$ and/or $\gamma + \delta > 1$, and $B_{0 \beta 1 0} = B_{0 \beta 0 1} = B_{0 0 \gamma \delta} = b_{\alpha \beta 0 0 } = 0$, and 

\begin{align}
t_{\rm{match}}^\omega \,(\eta,e,l) &= \sum_{\alpha, \beta, \gamma, \delta} C_{\alpha \beta \gamma \delta} \, e^\alpha \, \eta^\beta \, \cos( \gamma \, l \,+\, \delta \,e \, l \,+\, c_{\alpha \beta \gamma \delta})\,,
\label{eq:t_match_frequency}
\end{align}
for frequency, where $C_{\alpha \beta \gamma \delta} = c_{\alpha \beta \gamma \delta} = 0$ for $\alpha + \beta > 3 $ and/or $\alpha > 2$ and/or $\gamma + \delta > 1$, and $C_{0 \beta 1 0} = C_{0 \beta 0 1} = C_{0 0 \gamma \delta} = c_{\alpha \beta 0 0 } = 0$. The values for the coefficients $A_{\alpha \beta \gamma \delta}$, $B_{\alpha \beta \gamma \delta}$, $C_{\alpha \beta \gamma \delta}$, $a_{\alpha \beta \gamma \delta}$, $b_{\alpha \beta \gamma \delta}$, and $c_{\alpha \beta \gamma \delta}$ obtained by performing a fit to the numerical values are tabulated in Tables~\ref{tab:shift}-\ref{tab:freq} (see Appendix A).\\
\\
Figure~\ref{fig:amplitude_fit} shows comparisons between the numerically obtained values for $t_{\rm{shift}}$, $t_{\rm{match}}^\mathcal{A}$, and $t_{\rm{match}}^\omega$, with the values predicted by our analytical fits. 
The predictions are within $\pm 1M$ for numerical estimates for $t_{\rm shift}$, $t_{\rm{match}}^\mathcal{A}$ and $t_{\rm{match}}^\omega$. We show amplitude and frequency comparisons between the target hybrids and our models for three cases along with the full waveform for the $q=2$ case in Fig.~\ref{fig:td-model-q123-amp}.\\

\subsection{Validation}
\label{sec:Validation}

\begin{figure*}[htbp!]
\centering
    \includegraphics[width=0.32\textwidth]{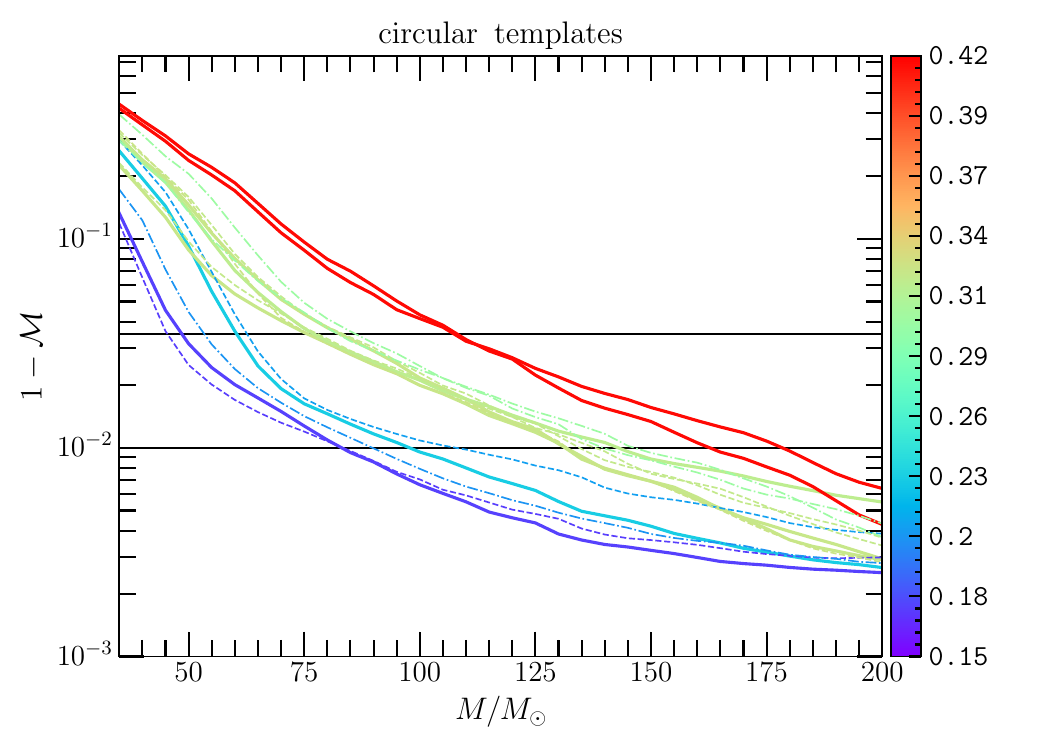}
    \includegraphics[width=0.32\textwidth]{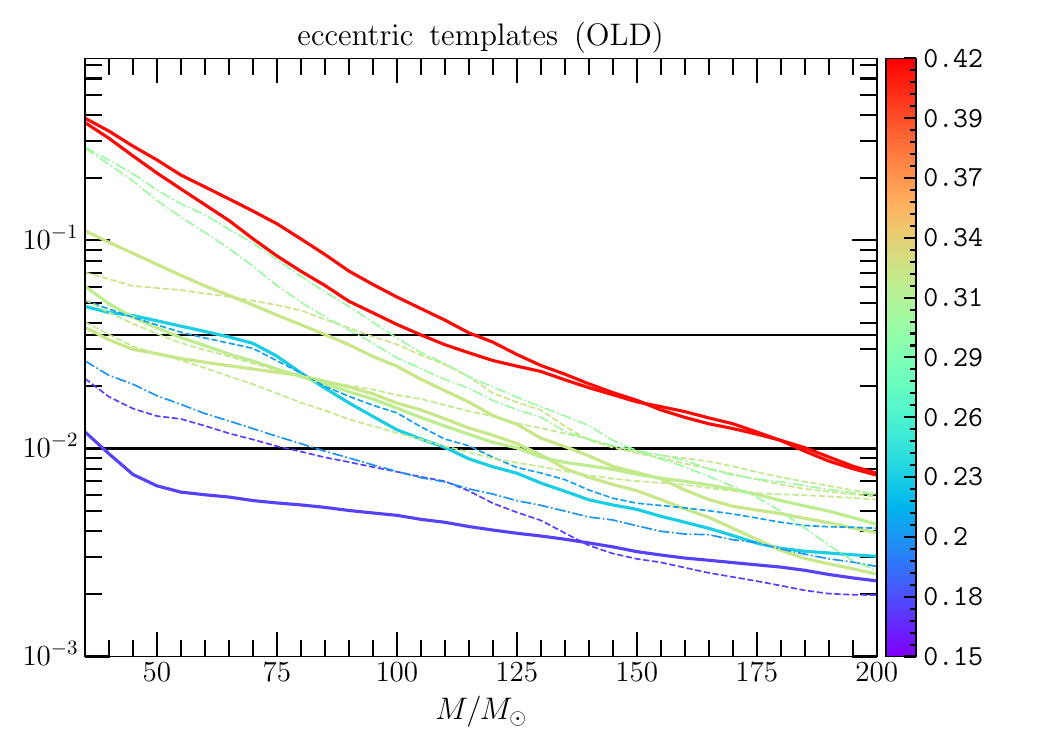}
    \includegraphics[width=0.32\textwidth]{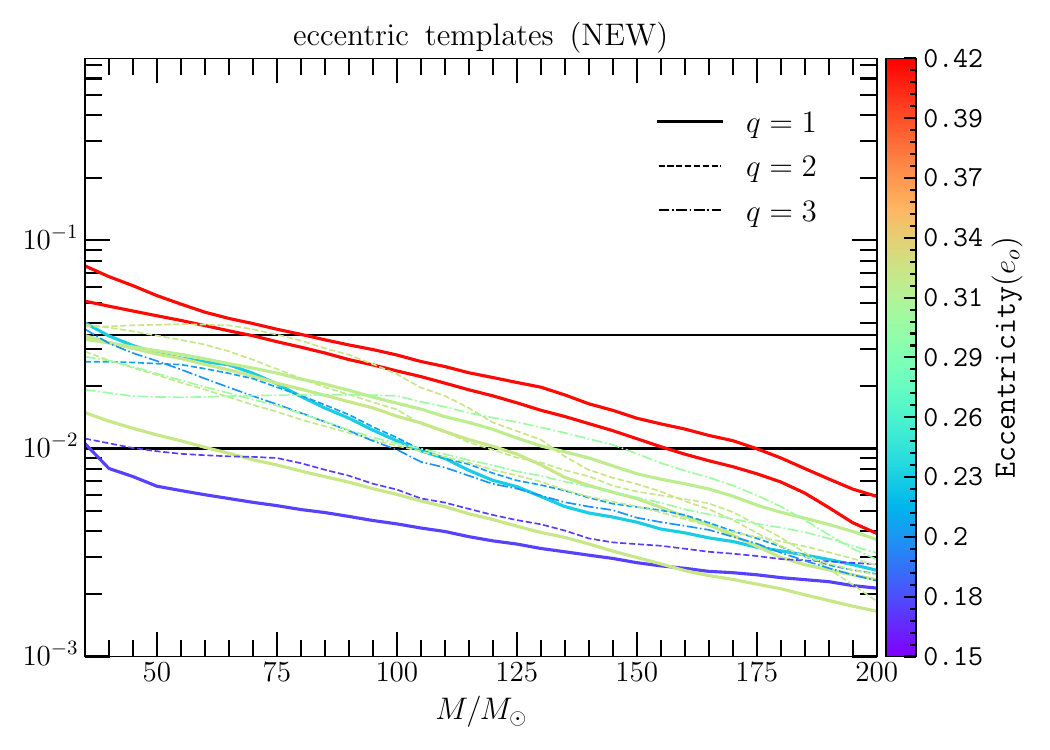}

    \caption{Mismatch as a function of total mass for hybrids (presented in Sec.~\ref{sec:hybrid_waveforms} and having initial eccentricity ($e_0$) within the calibration range of the model), when compared with the dominant mode ($\ell=2$, $|m|=2$) quasicircular templates (\textsc{SEOBNRv5}~\cite{Pompili:2023tna}; left panel) and eccentric templates [middle panel shows comparison with the model constructed in Paper I (old model) and right panel with the model constructed here (new model)]. The two horizontal lines indicate a match of 96.5\% (or mismatch of 3.5\%) and 99\% (or 1\% mismatch), respectively, and the color bar displays initial eccentricity ($e_0$) value at an initial GW frequency ($x_0$) for the respective target hybrid (see Table~\ref{tab:nrsims}).}
    \label{fig:match_ecc_temp_new}
\end{figure*}

\begin{figure*}[htbp!]
\centering
    \includegraphics[width=0.32\textwidth]{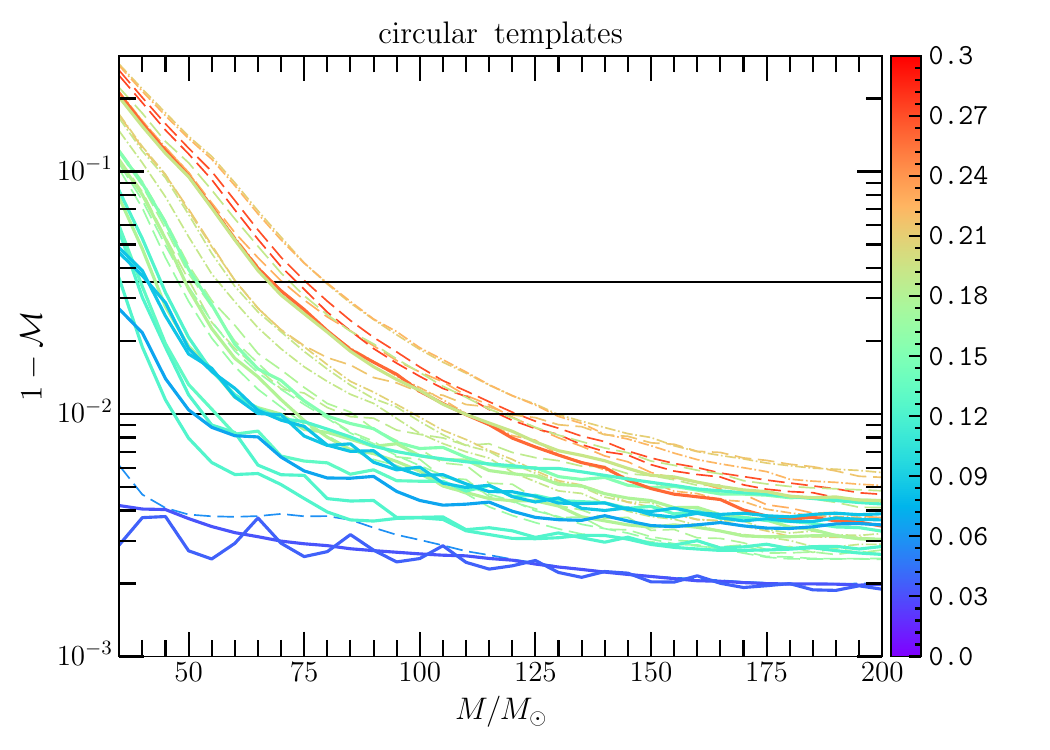}
    \includegraphics[width=0.32\textwidth]{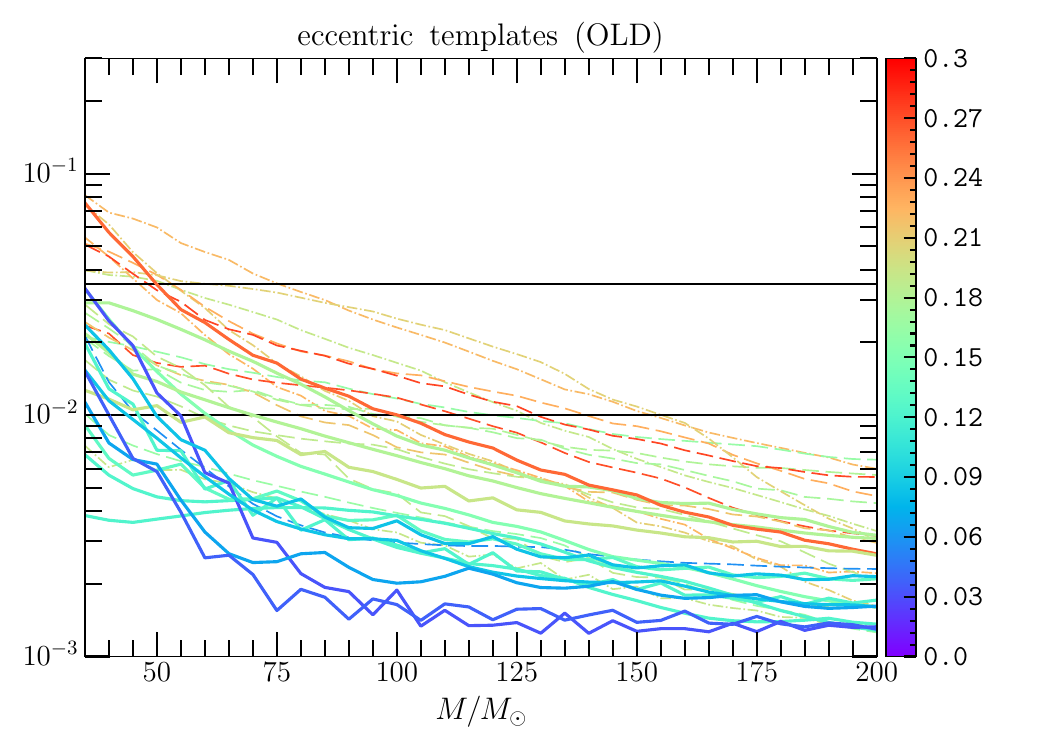}
    \includegraphics[width=0.32\textwidth]{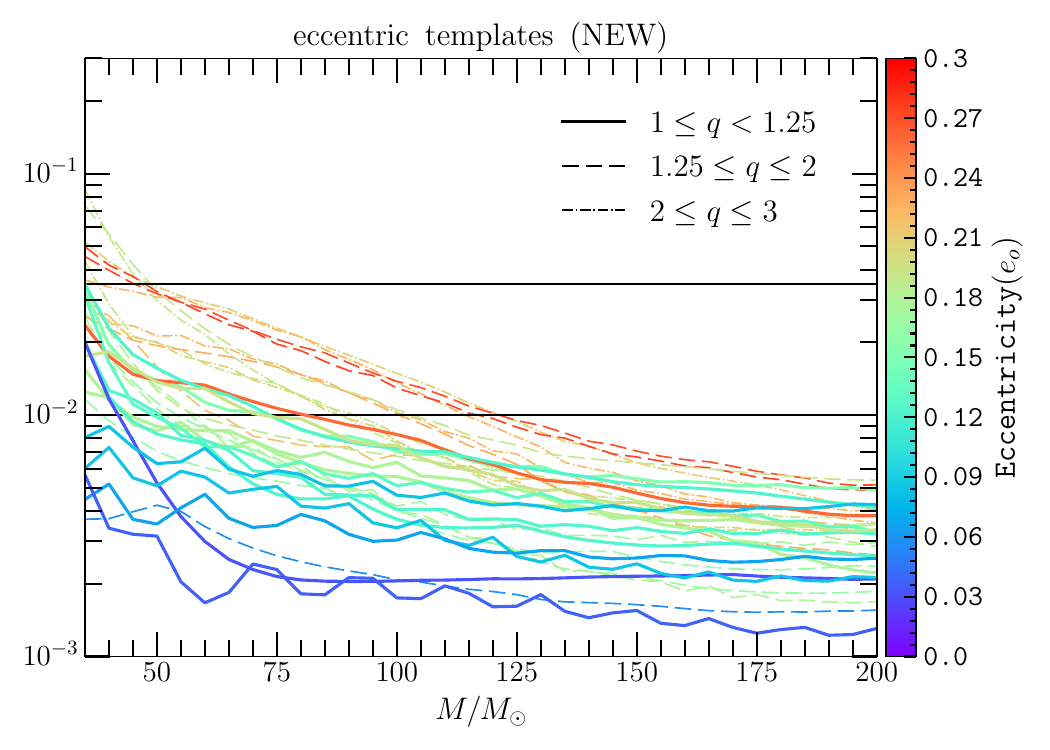}
    \caption{Same as Fig.~\ref{fig:match_ecc_temp_new} except mismatches are computed against an independent family of waveforms (\textsc{TEOBResumS-Dali}~\cite{Nagar:2024oyk}). Additionally, the parameter space explored by randomly sampling values for initial eccentricity ($e_0$) and mean anomaly ($l_0$) as well as of the mass ratio ($q$), in the range $0\lesssim e_0\lesssim0.3$, $-\pi\leq l_0\leq\pi$, and $1\lesssim q\lesssim3$, respectively.
    }
    \label{fig:match_ecc_22TEOB}
\end{figure*}

The performance of the model can be assessed from the plots against the training set hybrids presented in Fig.~\ref{fig:td-model-q123-amp}, as well as from the mismatch ($1-\mathcal{M}$) plots displayed in Fig.~\ref{fig:match_ecc_temp_new}. Note, however, that the hybrids used as targets in Fig.~\ref{fig:match_ecc_temp_new} are the ones that were constructed in Sec.~\ref{sec:hybrid_waveforms} and are different from the ones used in training the model, although, only the hybrids with $e_0$ values in the range sampled by training set hybrids (see Table~\ref{tab:nrsims}) are employed in validating the model. 
The model is optimized against these hybrids using the Nelder-Mead downhill simplex minimization algorithm of \textsc{Scipy} \cite{scipy2020}. This algorithm minimizes the mismatch between the target and the model over an initial set of three parameters -- ($e_0, l_0, f_0$). Certainly, the dominant mode model outperforms the quasicircular templates. On top of that, unlike Paper I, the current model provides $\geq96.5\%$ match against most of the hybrids for almost the entire range of parameters considered here (see right panel of Fig.~\ref{fig:match_ecc_temp_new}). For comparison, we also plot the mismatches of the model of Paper I optimized against the hybrids of the current work (middle panel of Fig.~\ref{fig:match_ecc_temp_new}). As can be seen there, our current model performs better than the one of Paper I, especially at the low mass end. For instance, matches for the old (new) model are larger than $96.5\%$ for systems heavier than $115M_\odot$ ($80M_\odot)$. The match degrades a little as we approach higher mass ratio as well as higher eccentricity cases. 
\\

Note that the target hybrids used in these mismatch computations 
include only the ($\ell=2, |m|=2$) mode so as to assess the actual performance of the dominant mode model. Given the quality of analytical fits it is expected that the analytical model performs well against the hybrids sharing the initial parameter values of the training set. 
Keeping this in mind, we also try to test our model against an independent waveform family \textsc{TEOBResumS-Dali}~\cite{Nagar:2024oyk}. 
For this comparison, we choose to sample a parameter space that is not identical to the training set hybrids. We choose to randomly sample 10,000 values each for reference initial eccentricity ($e_0$), mass ratio ($q$) and reference initial mean anomaly ($l_0$) in the range $0\lesssim e_0\lesssim0.3$, $1\lesssim q\lesssim3$ and $-\pi\leq l_0\leq\pi$ respectively; and the optimization is then performed for 32 sets. Note that the range for initial orbital eccentricity ($e_0$) is slightly different from the ones spanned by the hybrids. While the upper value is chosen conservatively to take $e_0=0.3$ to reduce any systematic differences between the model and the waveform \textsc{TEOBResumS-Dali}~\cite{Nagar:2024oyk} at large eccentricity values, near circular cases are also included to see if the model gradually produces the circular limit despite being trained on purely eccentric target models. \\

Next, the template (dominant eccentric model) is optimized against the target (\textsc{TEOBResumS-Dali}~\cite{Nagar:2024oyk}) using the same minimization algorithm used in validating the model against Sec.~\ref{sec:hybrid_waveforms} hybrids. The mismatch plot obtained is shown in Fig.~\ref{fig:match_ecc_22TEOB}. Additionally, for comparison, mismatches of \textsc{TEOBResumS-Dali}~\cite{Nagar:2024oyk} with quasicircular \textsc{SEOBNRv5}~\cite{Pompili:2023tna} templates and the model of Paper I are displayed in the left and middle panels respectively. Clearly, our current model seems to do better compared to the circular templates at the low mass end where the overlaps are $\gtrsim$ 96.5\% for nearly the entire range of parameters considered in the comparison. Compared to the model of Paper I, the current model again seems to perform better at the low mass end. For instance, matches for the old (new) model are larger than 96.5\% for systems heavier than 75$M_{\odot}$ (50$M_{\odot}$). The mismatches are comparable for heavier systems as expected since both the models assume circularized merger ringdown.

\section{An alternate model and inclusion of higher modes}
\label{sec:alt_model}
\subsection{Eccentric model based on TaylorT2 phase and PN corrected amplitudes}
\label{sec:TaylorT2 model}

\begin{figure*}[htbp!]
    \centering
    \includegraphics[width=0.49\textwidth]{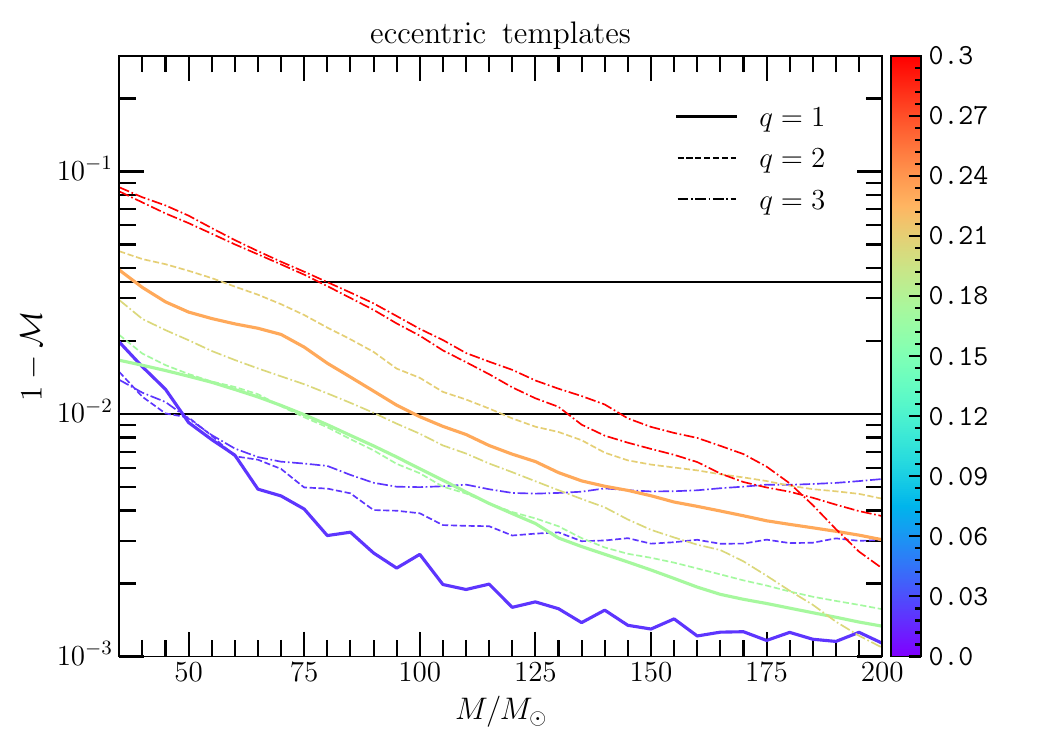}
    \includegraphics[width=0.49\textwidth]{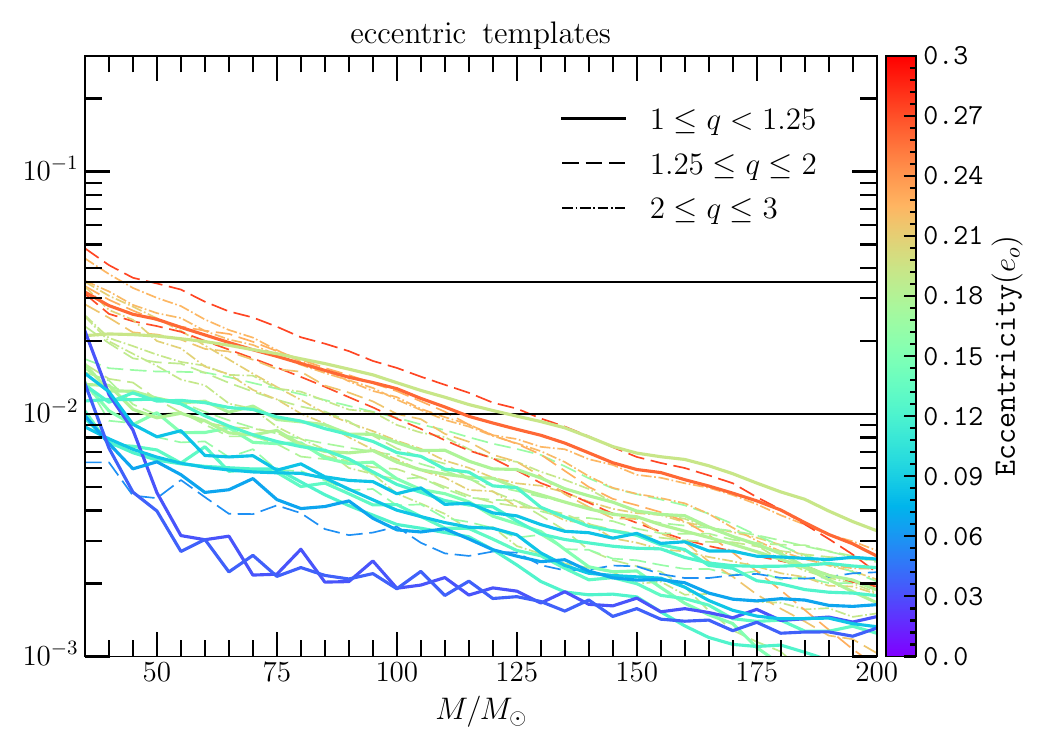} 
    \caption{Left: same as in Fig.~\ref{fig:match_ecc_temp_new} except, target waveforms (hybrids presented in Sec.~\ref{sec:hybrid_waveforms}) are compared with the alternate dominant mode model presented in Sec.~\ref{sec:TaylorT2 model}. Additionally, three circular hybrids (\textsc{HYB:SXS:BBH:1132}, \textsc{HYB:SXS:BBH:1167} and \textsc{HYB:SXS:BBH:1221}) from Paper I are also included to check the circular limit. Right: same as in Fig.~\ref{fig:match_ecc_22TEOB} except target waveforms (\textsc{TEOBResumS-Dali} \cite{Nagar:2024oyk}) are compared with the alternate dominant mode model presented in Sec.~\ref{sec:TaylorT2 model}.}  
    \label{fig:TT2_TEOB_optimized}
\end{figure*}

In this section, we discuss the possibility of finding a suitable alternative to the model constructed in the previous section without performing any additional calibration, i.e. the model fits for $t_{\rm{shift}}$, $t_{\rm{match}}^\mathcal{A}$ and $t_{\rm{match}}^\omega$ are the same. As mentioned earlier in the Sec.~\ref{sec:summary}, we propose to replace the PN model used in constructing the model in the previous section to include amplitude terms with higher PN accuracy in the model. We employ 3PN accurate expressions for the dominant mode amplitude of Refs.~\cite{Mishra:2015bqa,Boetzel:2019nfw,Ebersold:2019kdc} and a 3PN accurate phasing (based on \textsc{TaylorT2} approximant)~\cite{Moore:2016qxz} to construct this alternate model. Note that the inspiral part of the model presented in Sec.~\ref{sec:22-mode-model} was entirely based on the work of Ref.~\cite{Tanay:2016zog}.
The \textsc{EccentricTD} approximant is 2PN accurate in phase and only Newtonian accurate in amplitude for the eccentricity related effects, although is based on a superior (compared to \textsc{TaylorT2}) approximant, namely \textsc{TaylorT4} and should also work better for larger eccentricities as it includes corrections to sixth power in eccentricity, while the \textsc{TaylorT2} phase we use involves only leading order corrections of eccentricity although being 3PN accurate \cite{Moore:2016qxz}. 
To test the performance of this model, we compare it with hybrids constructed in Sec.~\ref{sec:hybrid_waveforms} having initial eccentricity $e_0 \leq 0.3$. Additionally, we include three circular hybrids (HYB:SXS:BBH:1132, 1167 and 1221) from Paper I to see if the model correctly reproduces the circular limit. The left panel of Fig.~\ref{fig:TT2_TEOB_optimized} shows this comparison for the dominant $(\ell=2, |m|=2)$ mode. We also compare our model with \textsc{TEOBResumS-Dali} \cite{Nagar:2024oyk} by computing overlaps on the same set of parameters used in validating the dominant mode model in Sec.~\ref{sec:Validation}. The results are shown in the right panel of Fig.~\ref{fig:TT2_TEOB_optimized}. Overlaps between the model and the target waveforms are better than $\sim96.5\%$ for almost the entire range considered here, making it a suitable alternative to the  \textsc{EccentricTD} model.\footnote{Note that, when the model is compared with target hybrids of Sec.~\ref{sec:hybrid_waveforms}, the overlaps are poorer ($\lesssim96.5\%$) for low total mass ($\lesssim80M_{\odot}$), high mass ratio ($q$=$3$) and high eccentricity ($e_0\sim0.3$) cases (see, left panel of Fig.~\ref{fig:TT2_TEOB_optimized}; see also, a discussion in Appendix~\ref{comparison of target wfs} illustrating the possible reason for poor mismatches.}

\subsection{Inclusion of higher order modes}
\label{sec:higher-mode-model}

\begin{figure*}[htbp!]
\centering
    \includegraphics[width=0.49\textwidth]{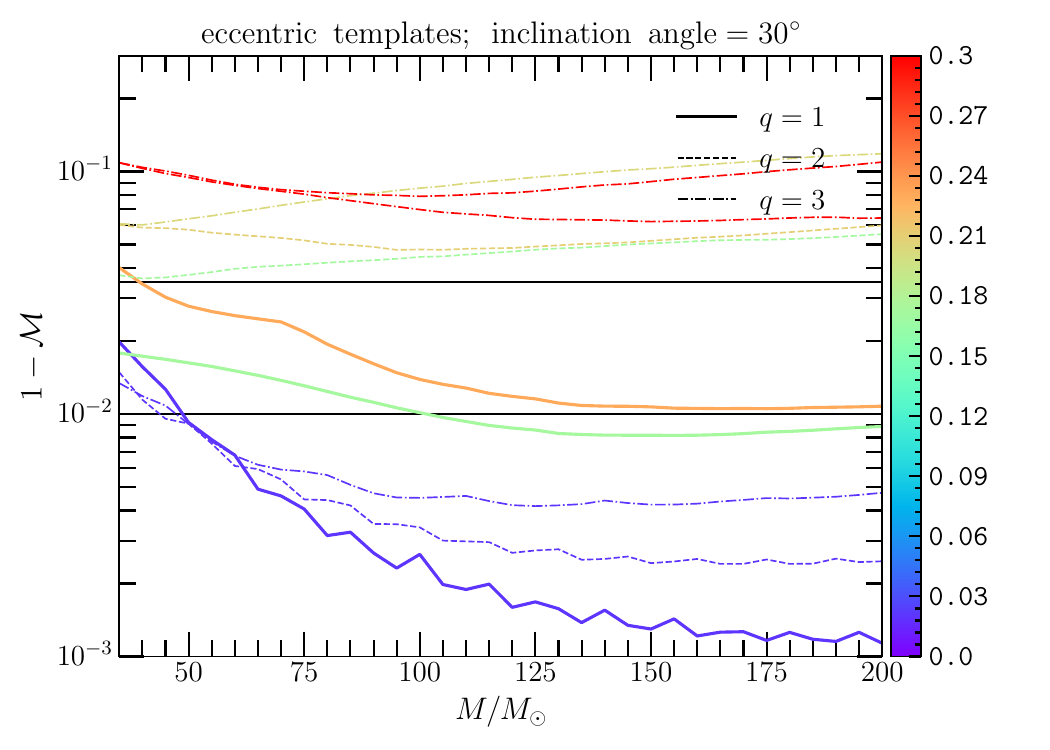}
    \includegraphics[width=0.49\textwidth]{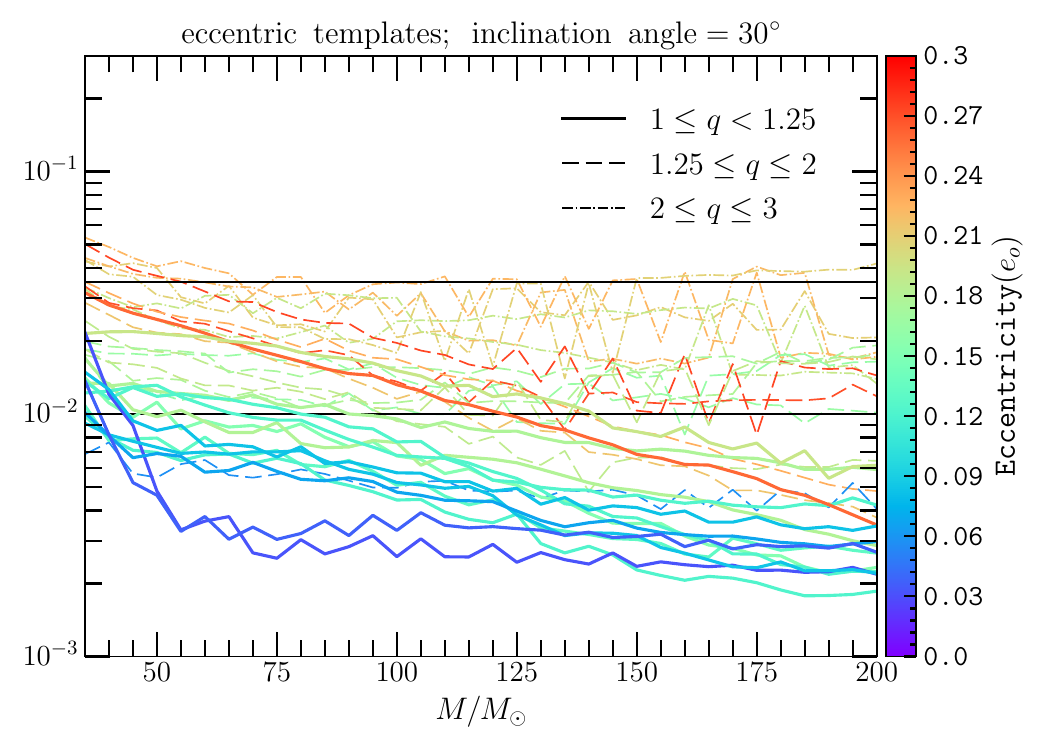}
   
    \caption{Similar to Fig.~\ref{fig:TT2_TEOB_optimized}, except that both the target and the model now include all modes included in our hybrids and the inclination angle of the binary is chosen to be $30^\circ$.}
    \label{fig:hm_model}
\end{figure*}

In this section, we extend the dominant mode model of Sec.~\ref{sec:TaylorT2 model} to obtain a higher mode model by including ($\ell$, $|m|$)=(2, 2), (2, 1), (3, 3), (3, 2), (4, 4), (4, 3) and (5, 5) modes. These are precisely the modes that are included in \textsc{SEOBNRv5HM} (Ref.~\cite{Pompili:2023tna}) which we use for the merger-ringdown part and also included in the hybrids constructed in Sec.~\ref{sec:hybrid_waveforms}. The (eccentric) inspiral and (quasicircular) merger-ringdown models are again matched using the analytical expressions for $t_{\rm{shift}}$, $t_{\rm{match}}^\mathcal{A}$ and $t_{\rm{match}}^\omega$, obtained for the dominant mode model in Sec.~\ref{sec:22-mode-model}, without performing any additional calibration for the individual higher modes (the higher mode model of Paper I was also obtained following the same strategy albeit, the HM model there only included the $\ell=|m|$ modes). The inspiral part of the model for each nonquadrupole mode is obtained by combining the mode amplitudes obtained in Refs.~\cite{Mishra:2015bqa, Boetzel:2019nfw, Ebersold:2019kdc} and orbital frequency [multiplied by an appropriate factor involving mode number; see Eq.~\eqref{eq:omega_lm}].  
The left panel of Fig.~\ref{fig:hm_model} compares our HM model with hybrids (also including HMs) presented in Sec.~\ref{sec:hybrid_waveforms}, for an inclination angle of $30^\circ$. Similar to Fig.~\ref{fig:TT2_TEOB_optimized} (left panel), here again we include the circular HM hybrids from Paper I. We find that the overlaps between our HM model and HM hybrids are poorer for cases with eccentricity values $\gtrsim0.2$ and mass ratios $\gtrsim2$; see also, Appendix~\ref{comparison of target wfs} for a related discussion. We also compare our HM model against the HM version of the waveform, \textsc{TEOBResumS-Dali}~\cite{Nagar:2024oyk} in the right panel of Fig.~\ref{fig:hm_model}. While all modes up to $\ell=8$ are included in the waveform polarizations implemented in \textsc{TEOBResumS-Dali}~\cite{Nagar:2024oyk}, we choose to include the same set of modes in the target and the model waveform to control the systematics. The orbital inclination angle is again chosen to be $30^\circ$. We find that the higher mode model recovers the \textsc{TEOBResumS-Dali} waveforms including HMs with accuracy better than $96.5\%$ for nearly the entire range of parameter values considered here. Note also, the strong oscillations in the right panel (comparison with \textsc{TEOBResumS-Dali} waveforms) of Fig.~\ref{fig:hm_model}. These oscillations seem to grow with increasing mass ratio but were absent in the right panel of Fig.~\ref{fig:TT2_TEOB_optimized} where only ($2, |2|$) modes were being compared, which is indicative of higher modes playing a role. Interestingly, these oscillations are absent in plots displaying the mismatches with hybrids (left panel of Fig.~\ref{fig:hm_model}). Larger mismatches with increasing mass (for nonequal mass cases) observed in Fig.~\ref{fig:hm_model} might be indicative of the fact that dominant mode fits (Eqs.~\ref{eq:t_shift_analytical}-\ref{eq:t_match_frequency}) are not optimal for constructing models involving nonquadrupole modes.

\subsection{Recovery of aligned spin binaries}
\label{sec:spins}
Although our alternate model (including higher modes) is nonspinning, we also test its performance against a spinning model for a few mildly spinning cases. 
We assume spin precession to be absent and thus binary's spin is described solely by the effective spin parameter, $\chi_{\rm eff}$. When expressed in terms of dimensionless spin components, $\chi_{i} = (\Vec{S_{i}} \cdot \hat{L})/m_{i}^2$, it reads~\cite{Divyajyoti:2023izl, Ajith:2009bn, Santamaria:2010yb}
\begin{equation}
    \chi_{\rm eff} = \frac{\chi_1 m_1 + \chi_2 m_2}{m_1+m_2}\,.
\end{equation}
Here, $m_{i}$ refers to the mass of the binary component having spin angular momentum $\Vec{S_{i}}$, and $\hat{L}$ denotes the unit vector along the direction of the orbital angular momentum of the binary.\\

\begin{figure}
    \centering
    \includegraphics[width=0.95\columnwidth]{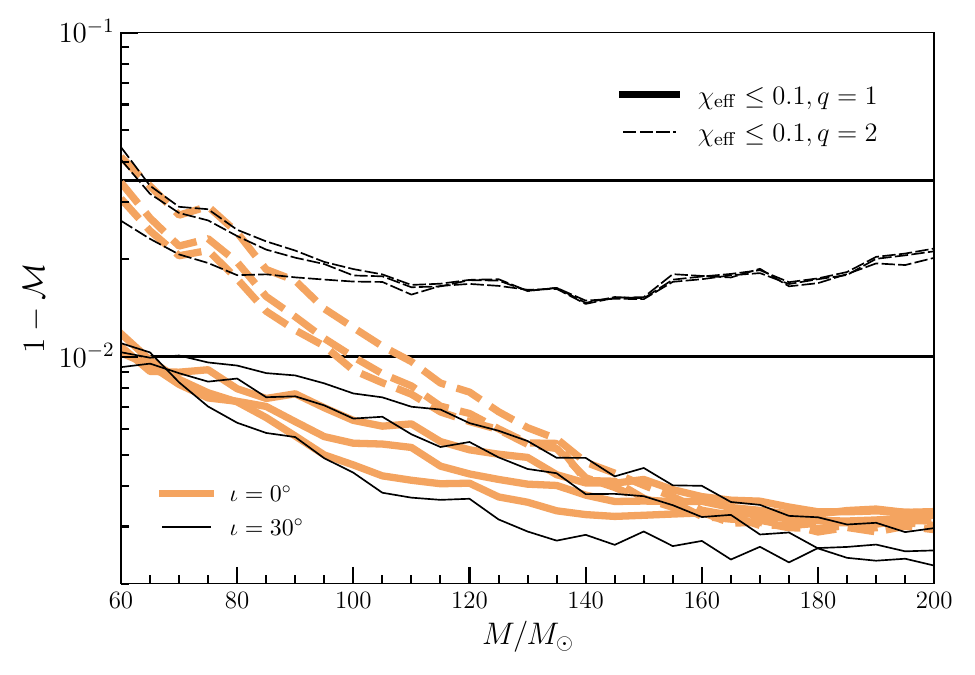}
    \caption{Recovery of mildly spinning target waveforms (\textsc{TEOBResumS-Dali}~\cite{Nagar:2024oyk}) with the alternate nonspinning model presented in Sec.~\ref{sec:TaylorT2 model} is displayed for two different mass ratio cases ($q=1$, 2), initial eccentricity ($e_0 = 0.1$ at $20$Hz) and a set of effective spin values (with $\chi_{\rm eff}\leq0.1$).}
    \label{fig:TT2_spinning}
\end{figure}

Figure~\ref{fig:TT2_spinning} compares our nonspinning model with the spinning version of \textsc{TEOBResumS-Dali}~\cite{Nagar:2024oyk} for initial eccentricity $e_0=0.1$ at 20 Hz. Thick curves represent the mismatch when the inclination angle ($\iota$) is set to zero 
for which only $|m|=2$ modes survive [in our case (2, 2) and (3, 2)]. 
Thin lines on the other hand, show mismatches for the inclination angle of $30^\circ$ and thus display mismatches with all $\ell=|m|$ and $\ell-1=|m|$ modes included in our higher mode model. For the equal mass case ($q=1$), we find excellent agreements with target models (matches $\gtrsim99\%$). For the $q=2$ case, the matches are still $\gtrsim 96.5\% $ for $\chi_{\rm{eff}} \leq 0.1$ for all systems with $M \gtrsim 60 M_\odot$. This clearly shows that despite being nonspinning in nature, the model could be used to analyze mildly spinning events observed routinely by current generation detectors such as LIGO and Virgo \cite{LIGOScientific:2018jsj, LIGOScientific:2020kqk, KAGRA:2021duu}. In other words, at least when analyzing mildly eccentric ($e_{\rm20Hz} \sim 0.1$), aligned spin systems with small effective spins ($\chi_{\rm eff}\leq0.1$) and masses ($\geq 60 M_\odot$), systematics due to neglect of spin effects may be ignored.

\section{Discussions and conclusions}
\label{sec:disc}
Our current work follows our earlier work of Ref.~\cite{Chattaraj:2022tay} (referred to as Paper I through preceding sections) where we developed a fully analytical dominant mode model for nonspinning binary black holes on elliptical orbits.  
We start by comparing 20 distinct NR simulations from SXS collaboration~\cite{Hinder:2017sxy} with a PN model obtained by combining inputs from Refs.~\cite{Ebersold:2019kdc, Boetzel:2017zza, Tanay:2016zog} for a set of spherical harmonic modes of gravitational waveforms. Figures~\ref{fig:pn_nr_comparison} and \ref{fig:22 mode comparison.} show this comparison. The two then are matched to obtain models we call hybrids and are used in training the model developed here. 
Details of model construction is discussed in Secs.~\ref{sec:numerical_model}-\ref{sec:analytical model}.
The model is then validated against the hybrids (presented in Sec.~\ref{sec:hybrid_waveforms}) that are not used in training the model.\footnote{As discussed earlier, hybrids with initial eccentricities outside the range of eccentricity values spanned by the training set, are not included when validating the model.} Our model reproduces the targets with matches better than $96.5\%$ for most of the parameter space (see right panel of Fig.~\ref{fig:match_ecc_temp_new}). 
We also validate our model against \textsc{TEOBResumS-Dali}~\cite{Nagar:2024oyk} for initial values of eccentricity ($e_0$), mass ratio ($q$), and mean anomaly ($l_0$) in the range $0\lesssim e_0\lesssim0.3$, $1\lesssim q\lesssim3$, and $-\pi\leq l_0\leq\pi$, respectively; the same can be considered as the range of validity for the model.   
Figure~\ref{fig:match_ecc_22TEOB} shows this comparison.\\

We also present an alternate model and its performance can be assessed from Fig.~\ref{fig:TT2_TEOB_optimized}. 
As can be seen, this alternate model performs approximately at the same level as the model presented in Sec.~\ref{sec:22-mode-model}, for the range of eccentricity values considered for validating the alternate model.  
Further, following Paper I, we extended this alternate model to include few leading $\ell=|m|$ and $\ell-1=|m|$ modes (up to $\ell=5$). 
Our HM model seems to reliably reproduce the target HM model (\textsc{TEOBResumS-Dali}~\cite{Nagar:2024oyk}) for inclination angles $\leq30^\circ$.
Finally, while our model(s) assumes spinless binary constituents, we also tried testing its suitability for analyzing signals with small spin magnitudes in the absence of spin precession. Figure~\ref{fig:TT2_spinning} compares our nonspinning model presented in Sec.~\ref{sec:TaylorT2 model}-\ref{sec:higher-mode-model} against the model of \textsc{TEOBResumS-Dali}~\cite{Nagar:2024oyk} (with spins switched on). Our nonspinning model is able to extract the equal mass spinning target waveforms with accuracy better than $99\%$ while  
matches larger than 96.5\% for the $q=2$ case. For higher spin magnitudes as well as higher mass ratios, mismatches are larger than 3.5\% in the low mass range, at least for unequal mass cases.\\

Note that, both the alternate model as well as the HM model were not obtained by calibrating against hybrids but rather we simply used the prescriptions presented in context of dominant mode model in Sec.~\ref{sec:analytical model} and thus could be improved; however, we restrict ourselves here to a proof of principle demonstration that alternate prescriptions for the dominant mode model as well as simple extensions like the one proposed here could be easily achieved and perform reliably. We leave such updates for a future work. 
Apart from being able to construct an improved model compared to the one presented in Paper I, in the current work we also address a few concerns with the model there. First, it was found that for nearly $10\%-15\%$ cases, the analytical fits for times for merger-ringdown attachment produced nonphysical values (beyond merger at $t>0$). Since the NR simulations used in this work essentially circularize by $30M$ before the merger \cite{Hinder:2017sxy}, we impose this condition for those nonphysical scenarios. This removes the discrepancy regarding the attachment times and ensures the validity of the model in the entire parameter space explored. 
Second, the model presented in Sec.~\ref{sec:TaylorT2 model} is significantly faster than the dominant mode model of Paper I. This is likely due to a speedup with the merger-ringdown model (\textsc{SEOBNRv5} \cite{Pompili:2023tna} instead of \textsc{SEOBNRv4} \cite{Bohe:2016gbl}). The waveform generation rate of \textsc{TaylorT2} model is $\sim 1.5$ times higher than the model based on \textsc{EccentricTD}. Moreover, when compared with the waveform \textsc{TEOBResumS-Dali}~\cite{Nagar:2024oyk}, these waveforms nearly have a $\sim 2$ times speedup and thus are likely to be useful for parameter estimation studies.\\

\section{Acknowledgments}
We thank Kaushik Paul and Prayush Kumar for sharing invaluable insights in validating our model. 
We thank the authors of Ref.~\cite{Nagar:2024oyk} for making the implementation of the \textsc{TEOBResumS-Dali} available for public use \cite{teobresums2024}, and Divyajyoti for helping us with technical details of the implementation and waveform generation. We thank Md Arif Shaikh for useful comments and suggestions on our manuscript. T.R.C. acknowledges the support of the National Science Foundation Award PHY-2207728. C.K.M. acknowledges the support of SERB's Core Research Grant No.~CRG/2022/007959.

This document has LIGO preprint number LIGO-P2400355. 

\section{Data Availability}
No data were created or analyzed in this study.

\appendix
\section{Coefficients of the analytical fit}
Here we tabulate the coefficients of the analytical fits for the parameters $t_{\rm shift},t_{\rm match}^\mathcal{A}$ and $t_{\rm match}^\omega$ obtained in Sec~\ref{sec:analytical model}. The expressions are given in Eqs.~\eqref{eq:t_shift_analytical}, \eqref{eq:t_match_amplitude} and \eqref{eq:t_match_frequency}.

\vspace*{0.2cm}
\begin{table}[h!]
        \centering
        \begin{tabular}{p{0.15\columnwidth}C{0.25\columnwidth}C{0.3\columnwidth}C{0.235\columnwidth}}
            \hline
            \hline
            $A_{\alpha \beta 0 0}$ & $\beta = 0$ & 1 & 2  \\
\hline
$\alpha = 0$ & $-622.279$   & $-330.308$  & $-27.0717$   \\
1            & $-141.883$    & $6787.92$  & $0$  \\
2            & $5132.37$   & $-17060.5$   & $0$   

        \end{tabular}
        
        \vspace*{-0.2cm}
        
        \centering 
        \begin{tabular}{p{0.15\columnwidth}C{0.55\columnwidth}C{0.25\columnwidth}}
            \hline
            \hline
            $\alpha \beta \gamma \delta$ & $A_{\alpha \beta \gamma \delta}$ & $a_{\alpha \beta \gamma \delta}$\\
\hline
$1010$ & $-2030.2$    & $-5.38518$  \\
$1101$ & $250.49$     & $1.04693$  \\
$2010$ & $6608.44$     & $-18.0212$ \\ 
\hline
\hline
            % \hline
        \end{tabular}

        \caption{Table of coefficients for the analytical expression of $t_{\rm{shift}}$ in Eq.~\eqref{eq:t_shift_analytical}. All other coefficients not included in the table are zero. }
        \label{tab:shift}
\end{table}

\begin{table}[h!]
        \centering
        \begin{tabular}{p{0.15\columnwidth}C{0.25\columnwidth}C{0.3\columnwidth}C{0.235\columnwidth}}
            \hline
            \hline
            $B_{\alpha \beta 0 0}$ & $\beta = 0$ & 1 & 2  \\
\hline
$\alpha = 0$ & $24215.3$              & $-120939.0$             & $91243.2$              \\
1            & $-166056.$             & $673029.$            & $0$   \\
2            & $328096.$              & $-1.34817 \times 10^6$            & $0$ 

        \end{tabular}

        \vspace*{-0.2cm}
        
        \centering 
        \begin{tabular}{p{0.15\columnwidth}C{0.55\columnwidth}C{0.25\columnwidth}}
            \hline
            \hline
            $\alpha \beta \gamma \delta$ & $B_{\alpha \beta \gamma \delta}$ & $b_{\alpha \beta \gamma \delta}$\\
\hline
$1010$ & $-1033.39$    & $119.166$  \\
$1110$ & $3251.62$   & $-980.368$ \\
$2001$ & $-5624.0$    & $681.809$ \\
\hline
\hline 

        \end{tabular}

        \caption{Table of coefficients for the analytical expression of amplitude $t_{\rm{match}}$ in Eq.~\eqref{eq:t_match_amplitude}. All other coefficients not included in the table are zero. }
        \label{tab:amplitude}
\end{table}

\begin{table}[h!]
        \centering
        \begin{tabular}{p{0.1\columnwidth}C{0.28\columnwidth}C{0.3\columnwidth}C{0.25\columnwidth}}
            \hline
            \hline
            $C_{\alpha \beta 0 0}$ & $\beta = 0$ & 1 & 2  \\
\hline 
$\alpha = 0$ & $-1.87711\times 10^6$                & $8.92879 \times 10^6$   & $-5.75326 \times 10^6$   \\
1            & $1.30607 \times 10^7$   & $-5.21053 \times 10^7$    & $0$  \\
2            & $-2.51977 \times 10^7$    & $1.04979 \times 10^8$   & $0$                         

        \end{tabular}

        \vspace*{-0.2cm}
        
        \centering 
        \begin{tabular}{p{0.15\columnwidth}C{0.55\columnwidth}C{0.25\columnwidth}}
            \hline
            \hline
            $\alpha \beta \gamma \delta$ & $C_{\alpha \beta \gamma \delta}$ & $c_{\alpha \beta \gamma \delta}$\\
\hline
$1110$ & $764721.0$             & $290.153$ \\
$1010$ & $181822.0$   & $61.2714$ \\
$2001$ & $-1.30815 \times 10^6$   & $446.125$ \\
\hline
\hline

        \end{tabular}

        \caption{Table of coefficients for the analytical expression of frequency $t_{\rm{match}}$ in Eq.~\eqref{eq:t_match_frequency}. All other coefficients not included in the table are zero. }
        \label{tab:freq}
\end{table}
%%%%%%%%%%%%%%%%%%%%%%%%%%

%%%%%%%%%%%%%%%%%%%%%%%%%%%%%%%%%%%%%%%%%%%%
\pagebreak
\section{Validation against imperfect targets and related systematics}
\label{comparison of target wfs}
While we choose to validate the model against a set of hybrids and the \textsc{TEOBResumS-Dali} waveforms~\cite{Nagar:2024oyk}, these target models may contain small modeling errors. In this section, we provide a simplistic measure of such errors. Let, $\Delta$ be the error in a model that can be quantified by comparing the model with a target waveform. If we denote the hybrids waveforms as model $A$ (say for now, the perfect target), \textsc{TEOBResumS-Dali} waveforms as model $B$ (an imperfect target) and TaylorT2-based model(s) of Sec.~\ref{sec:alt_model} as model $C$, then $|\Delta(A,B)|$ and $|\Delta(A,C)|$ will represent the modeling error in $B$ and $C$ assuming $A$ as a target model. On the other hand, $|\Delta(B,C)|$ represents the error in model $C$ assuming $B$ as target model, although, it has an error represented by $|\Delta(A,B)|$ and thus an upper limit on error in $C$ can be given by the sum, $|\Delta(A,B)|+|\Delta(B,C)|$. We test this simplistic upper limit with an example. We compute the mismatch (say quantifying $\Delta$) between model $A$, $B$, and $C$ assuming a total mass of 35$M_{\odot}$, $q=3$ and $e_0=0.3$. For the dominant ($\ell=2,|m|=2$) mode, we find that both $|\Delta(A,B)|$ and $|\Delta(B,C)|$ are $\sim 5\%$. This implies $|\Delta(A,C)|\leq|\Delta(A,B)|+|\Delta(B,C)|\sim 10\%$. Similarly, for the model including HMs, we find that $|\Delta(A,B)|\sim 6\%,\ |\Delta(B,C)|\sim 5\%$ or the modeling error,  $|\Delta(A,C)|\leq|\Delta(A,B)|+|\Delta(B,C)|\sim11\%$. As can also be verified from the left plots of Fig.~\ref{fig:TT2_TEOB_optimized} and ~\ref{fig:hm_model}, for such a system, the mismatch between the dominant model based on TaylorT2 (model $C$), and its HM version with target hybrids (model $A$) is $\sim 9\%$ and $\sim 10\%$, respectively. This naturally explains the poor match between \textsc{TaylorT2} Model and the hybrids shown in the left panel of Fig.~\ref{fig:TT2_TEOB_optimized} and ~\ref{fig:hm_model}. In particular, the mismatches seem to grow beyond the tolerance ($\sim3.5\%$) for large eccentricity ($e_0\gtrsim0.2$) and large mass ratio ($q\gtrsim2$) cases. On the other hand, the mismatch between our model and \textsc{TEOBResumS-Dali} waveforms seem to be within the tolerance. This hints at possible modeling errors in our target models and appropriate caution is called for when using the models.

%%%%%%%%%%%%%%%%%%%%%%%%%%%%%%%%%%%%%%%%%%%%%%%%%%%%

%\pagebreak

\bibliographystyle{apsrev4-1}
\bibliography{Manna_et_al_2024}
\end{document}